\documentclass[sigconf,screen, nonacm]{acmart}

\usepackage{graphicx}
\usepackage{subcaption}
\usepackage{xcolor}
\usepackage{multirow}
\usepackage{listings}

\newcommand{\rapper}[1]{\textcolor{purple}{rapper: #1}}
\newcommand{\nikos}[1]{\textcolor{blue}{nikos: #1}}
\newcommand{\panos}[1]{[\textcolor{magenta}{Panos: #1}]}

\newcommand{\eat}[1]{}

\newcommand{\stitle}[1]{\vspace{1ex}\noindent\textbf{#1}}

\definecolor{keyword}{HTML}{2771a3}
\definecolor{pattern}{HTML}{b53c2f}
\definecolor{string}{HTML}{be681c}
\definecolor{relation}{HTML}{7e4894}
\definecolor{variable}{HTML}{107762}
\definecolor{comment}{HTML}{8d9094}
\begin{document}

% \setlength{\textfloatsep}{10pt plus 2pt minus 4pt}
% \setlength{\floatsep}{10pt plus 2pt minus 4pt}

%% Your Title (Remember the [experiments] tag for short papers)
\title{Indexicon: A Spatial Indexing Library}

%% Author Information (Comment this out or change documentclass to anonymous if double-blind)
\author{Panagiotis Simatis}
\affiliation{%
  \institution{Department of CS and Engineering}
  \institution{University of Ioannina}
  \city{Ioannina}
  \country{Greece}
}
\email{p.simatis@uoi.gr}

\author{Panagiotis Bouros}
\affiliation{%
  \institution{Institute of Computer Science}\institution{Johannes Gutenberg University Mainz}
  \city{Mainz}
  \country{Germany}
}
\email{bouros@uni-mainz.de}

\author{Nikos Mamoulis}
\affiliation{%
  \institution{Department of CS and Engineering}
  \institution{University of Ioannina}
  \city{Ioannina}
  \country{Greece}
}
\email{nikos@cs.uoi.gr}

\renewcommand{\shortauthors}{Simatis et al.}

%% Abstract
\begin{abstract}
Spatial indexing is foundational to Geographic Information Systems (GIS) and multi-dimensional data management, yet the current open-source landscape poses a significant barrier to research that employs or benchmarks spatial access methods.
We observe that most of the existing open-source libraries include a single index. Some are hindered by complex dependencies, missing critical functionalities, inconsistent APIs, and rigid constraints regarding the support of spatial data types. To address this issue, we introduce \mbox{\textbf{Indexicon}}: a unified, highly portable, extendable, open-source spatial indexing library, designed specifically for rapid integration and ease of modification of main-memory spatial access methods. Indexicon provides a comprehensive suite of popular tree-based spatial access methods, including the R-tree, Quad-tree variants, and the KD-tree.
Each structure is meticulously implemented as a self-contained, single-file, header-only C++ template with zero external dependencies beyond the standard library. Crucially, every index features a uniform interface, natively supporting bulk-loading, dynamic insertions/deletions, range queries, $k$-nearest neighbor ($k$NN) search, and structural statistics tracking. We also present an extensive performance evaluation of Indexicon against well-established and widely used implementations of these structures (including Boost Geometry, PCL, and Nanoflann) across six real-world geographic datasets. Our results demonstrate that Indexicon's lightweight design matches or outperforms existing state-of-the-art implementations while offering unmatched architectural flexibility. To foster reproducible spatial research, we open-source the complete library alongside our datasets and query workloads.
\end{abstract}

% %% CCS Concepts and Keywords (Generate your real ones at http://dl.acm.org/ccs.cfm)
% \begin{CCSXML}
% <ccs2012>
%    <concept>
%        <concept_id>10002951.10002952.10002971.10003451</concept_id>
%        <concept_desc>Information systems~Data layout</concept_desc>
%        <concept_significance>500</concept_significance>
%        </concept>
%  </ccs2012>
% \end{CCSXML}

% \ccsdesc[500]{Information systems~Data layout}

\keywords{Spatial indexing, access methods, data structures, R-tree, Quad-tree, KD-tree, benchmarking}

\maketitle

%% --- INTRODCTION ---

\section{Introduction}
\label{sec:intro}

Spatial computing, geographic information systems (GIS), and modern spatiotemporal architectures rely fundamentally on efficient access methods. Multi-dimensional data structures such as the Quad-tree \cite{DBLP:journals/acta/FinkelB74, DBLP:books/daglib/0014978, DBLP:journals/acta/OvermarsL82}, the KD-tree \cite{DBLP:journals/cacm/Bentley75} and the R-tree \cite{DBLP:conf/sigmod/Guttman84, DBLP:conf/sigmod/BeckmannKSS90}, have served as the bedrock of spatial data management for several decades. These indices drastically prune the search space for popular retrieval operations, such as range and nearest neighbor queries and extensions thereof \cite{DBLP:series/synthesis/2011Mamoulis}. Besides, proposed solutions for a plethora of data analysis problems, e.g., in 
computer vision and image retrieval \cite{DBLP:conf/sigmod/KatayamaS97,ROBINSON1996209,DBLP:conf/mmm/WuHNH11,DBLP:conf/cvpr/Silpa-AnanH08,DBLP:conf/dasfaa/OhFKMB01}, 
top-$k$ and ranking search \cite{DBLP:conf/vldb/MouratidisYPM06,DBLP:conf/sigmod/VlachouDNK13,muja_flann_2009,DBLP:journals/pami/MujaL14}, 
skyline computation  \cite{DBLP:conf/sigmod/PapadiasTFS03,DBLP:conf/icde/MorsePG06,DBLP:conf/icde/ParkKPKI09,DBLP:conf/vldb/DellisS07,DBLP:journals/jksucis/SongWGWW25,DBLP:conf/sigmod/LianC08},
clustering \cite{DBLP:conf/kdd/EsterKSX96,DBLP:conf/sigmod/AnkerstBKS99,DBLP:conf/vldb/EsterKSWX98,DBLP:journals/topc/ChenRB25,DBLP:journals/jpdc/WuCTW22,DBLP:journals/pami/KanungoMNPSW02,DBLP:conf/kdd/PellegM99}, temporal  \cite{DBLP:journals/pacmmod/ChristodoulouBM24,DBLP:journals/vldb/SimatisCBM26} and
uncertain data management \cite{DBLP:journals/tods/TaoXC07,DBLP:journals/isci/YongLKH14,DBLP:conf/icde/BerneckerEKMRZ11,DBLP:journals/tkde/ParkerIGS09,DBLP:journals/tkde/LiuYYL13},
among others, rely on and extend spatial indices.

Two geometry types are typically indexed by such data structures: points and minimum bounding boxes (MBBs). Extended geometries, such as polygons, are approximated by MBBs and the spatial index organizes them and processes the {\em filter} step of the spatial query, which retrieves the objects whose MBB satisfies the query predicate. In a {\em refinement} step, the exact geometries of these objects are fetched and verified \cite{DBLP:conf/icde/BrinkhoffKS93}.

In the past, several surveys \cite{DBLP:journals/pvldb/ChenGZJYY17,DBLP:journals/tkde/LiWDLCGCCLLC24,DBLP:journals/vldb/LiuLZSC25,DBLP:journals/csur/GaedeG98,DBLP:journals/dase/PandeyRKK21,DBLP:journals/pvldb/PandeyKNK18} have investigated the performance and the scalability of spatial libraries and multi-dimensional (or spatial) indexing under different settings.
However, as (geo)spatial information expands in velocity and applications, ranging from LiDAR point clouds to real-time GPS tracking and from medical to scientific data, the engineering implementation details of spatial structures have become as critical as their achieved query throughput and their theoretical complexity bounds. 
%As the memory capacities of modern commodity hardware increases, the main performance objective when using spatial access methods shifts from I/O latency to computational cost. \panos{I believe ``shifts from I/O cost to memory latency and computational cost'' is better}
Essentially, while these benchmarks and surveys offer invaluable insights, they fail to delve into the technical and software engineering challenges of spatial indexing.
% predominantly test fragmented, pre-existing software implementations burdened by opaque codebases and incomplete or incompatible APIs.

Researchers and practitioners in the spatial database community face a highly fragmented and rigid open-source software ecosystem. Modern research frequently demands modifying an index to handle non-traditional tasks, specialized hardware acceleration layouts, or non-native query predicates. However, available open-source implementations are typically single-index focused, burdened with deeply nested multi-file dependencies, and incompatible APIs. Furthermore, many libraries omit critical operational capabilities like dynamic deletions, robust bulk-loading routines, or structural statistics reporting. This systemic fragmentation not only prevents true apples-to-apples performance comparisons between data structures in customized search tasks, but also drastically elevates the barrier to entry for developing and testing novel 
multi-dimensional search algorithms and extensions of spatial data structures.
%spatial indexing variants and search algorithms.
%
In view of the above, we define the following key challenges in designing and developing libraries for spatial indexing:
\begin{enumerate}
% [noitemsep,topsep=2pt,parsep=0pt,partopsep=0pt,leftmargin=0.5cm]
    \item \textbf{Features and capabilities}. An ideal library includes implementations of multiple spatial data structures for indexing different data types (e.g., points and minimum bounding boxes MBB) in both the 2D and 3D space. The library should also support construction and maintenance operations, and offer evaluation methods for different query types.
    \item \textbf{Code complexity and extendability}. The spatial index implementations should exhibit low structural and code complexity, allowing researchers and practitioners to rapidly incorporate them into their projects. At the same time, the cost of modifying and adapting the codebase should be low, allowing users to extend the functionality and the feature set of the library, e.g., to design a new index building upon existing ones or to develop and test unsupported (potentially novel) search tasks.
    \item \textbf{Performance}. For each included spatial data structure, the library should implement the recommended storage layout and the state-of-the-art index construction and search algorithms in the literature. In addition, it should allow the user to optimize performance by tuning the index parameters (e.g., node capacity), based on the data characteristics and application requirements.
\end{enumerate}

As we discuss in Section~\ref{sec:existing}, no existing library or open-source implementation of spatial indexing manages to successfully address all three challenges.
To address this,
% \vspace{4ex}
% \noindent\panos{checkpoint}
% \stitle{Contributions}.
% In view of the above, 
% this paper systematically evaluates the architectural design, and real-world performance of foundational spatial indexing methods. We frame our work around two key challenges:
% \begin{enumerate}
% % [noitemsep,topsep=2pt,parsep=0pt,partopsep=0pt,leftmargin=0.5cm]
%     \item \textbf{Flexibility}: Can we develop implementations of classic indices that simultaneously reduce structural code complexity and maximize adaptability for unsupported spatial tasks?
%     \item \textbf{Performance}: Can our developed spatial indices match or even outperfom current and widely used implementations across various configurations, datasets, and query workloads?
% \end{enumerate}
% In the course of addressing these challenges, 
we have developed and we constantly evolve \textbf{Indexicon}\footnote{The name \emph{Indexicon} is a lighthearted nod to the classic fictional grimoires, framing the library as a definitive tome of spatial indexing structures.}, a comprehensive spatial indexing library and benchmarking suite tailored for modern main-memory applications. Indexicon is engineered from the ground up to empower researchers and practitioners with a portable, fully open-source, and easily modified development environment. Concretely, our work delivers several contributions:
\begin{enumerate}
% [noitemsep,topsep=2pt,parsep=0pt,partopsep=0pt,leftmargin=0.5cm]
    \item \textbf{A portable architecture:} We consolidate several spatial indexing structures into a single cohesive, header-only C++ framework. By eliminating API discrepancies and requiring zero external dependencies beyond the standard library, Indexicon provides researchers with a highly portable, drop-in development environment.
    \item \textbf{A standardized benchmarking suite:} To foster reproducible spatial research, we curate a robust evaluation sandbox featuring several real-world geographic datasets that include points or minimum bounding boxes (MBBs) of geometries,
    % of distinct structural profiles (e.g., points, MBBs),
    while offering custom query generation functionality.
    \item \textbf{A comprehensive empirical evaluation:} We present a comprehensive performance study, comparing Indexicon against data structures from widely used spatial libraries, including Boost Geometry, GEOS, PCL, and Nanoflann. By evaluating 
    % these libraries
    under identical hardware constraints, data types, and workloads, we eliminate historical benchmarking asymmetries and establish a fair performance baseline.
% \panos{need to clarify/remind here that we compare data structures not the entire libraries}
    \item \textbf{Performance advantage}: Our tests reveal that Indexicon successfully matches or even exceeds the ingestion speed and efficiency of popular baselines in core range and $k$NN search operations, while providing unprecedented architectural flexibility. 
\end{enumerate}

\stitle{Outline}. The remainder of this paper is structured as follows. Section~\ref{sec:existing} overviews the landscape of popular spatial indexing libraries and implementations. Section~\ref{sec:lib} introduces the design philosophy and internal engineering of the Indexicon library. Section \ref{sec:study} presents our experimental analysis\eat{details our rigorous experimental setup, parameter tuning, and extensive comparative evaluations}. 
%Section \ref{sec:related} surveys related indexing frameworks, and 
Finally, Section \ref{sec:conclusion} concludes the paper with directions for future work.

%% --- Indexicon ---

\section{Spatial Indexing Landscape}
\label{sec:existing}

\begin{table*}[ht]
\caption{Overview of popular spatial indexing libraries and implementations}
\label{tab:index_capabilities}
\centering
% \scriptsize
\resizebox{\textwidth}{!}{%
\begin{tabular}{lcccccccc}
\toprule
\textbf{Indices}    &\textbf{Dimensionality}    &\textbf{Data}      &\textbf{Bulk}      &\textbf{Maintenance}   &\textbf{Range}     &\textbf{$k$NN}     &\textbf{Code}          &\textbf{Extensibility}\\
                    &                           &\textbf{types}     &\textbf{loading}   &                       &\textbf{queries}   &\textbf{queries}   &\textbf{complexity}    &\\
\midrule
\multicolumn{7}{l}{\textbf{Indexicon} (this paper)} \\
% \midrule
R-tree              &Any                        &Point/MBB    &\checkmark         &Insert/Delete          &\checkmark         &\checkmark         &Low                    &High\\
Quad-tree           &2D                         &Point/MBB    &\checkmark         &Insert/Delete          &\checkmark         &\checkmark         &Low                    &High\\
Oct-tree            &3D                         &Point              &\checkmark         &Insert/Delete          &\checkmark         &\checkmark         &Low                    &High\\
KD-tree             &Any                        &Point              &\checkmark         &Insert/Delete          &\checkmark         &\checkmark         &Low                    &High\\
\midrule
\multicolumn{7}{l}{\textbf{Boost Geometry} \cite{boost}} \\
% \midrule
R-tree              &Any                        &Point/MBB    &\checkmark         &Insert/Delete          &\checkmark         &\checkmark         &High                   &Low\\
\midrule
\multicolumn{7}{l}{\textbf{CGAL} \cite{cgal:project}} \\
Orth-tree           &Any                        &Point              &\checkmark         &-                      &Approximate      &\checkmark           &High                   &Low\\
KD-tree             &Any                        &Point              &\checkmark         & Insert (re-build required)/Delete (lazy)   &\checkmark         &\checkmark         &High                   &Low\\
\midrule
\multicolumn{7}{l}{\textbf{GEOS } \cite{geos}} \\
R-tree              &2D                         &Point/MBB    &\checkmark         &-                      &\checkmark         &Only for $k=1$     &High                 &Low\\
Quad-tree           &2D                         &Point/MBB    &-                  &Insert/Delete          & Approximate       &-                  &High                 &Low\\
\midrule
\textbf{LibSpatialIndex} \cite{libspatialindex}\\
R-tree              &Any                        &Point/MBB    &\checkmark         &Insert/Delete          &\checkmark         &\checkmark         &High                   &Low\\
\midrule
\textbf{PCL} \cite{pcl} \\
Oct-tree            &3D                         &Point              &-                  &Insert                 &Float-based        &Float-based         &High                 &Low\\
KD-tree             &3D                         &Point              &\checkmark         &-                      &Float-based radius &Float-based         &High                 &Low\\
\midrule
\textbf{Nanoflann} \cite{nanoflann}\\
KD-tree             &Any                        &Point              &\checkmark          &Insert/Delete (lazy) &Static only/Dynamic only radius         &\checkmark         &Medium                    &High\\
\midrule
\textbf{Vigier} \cite{pvigier} \\
Quad-tree           &2D                         &Point/MBB    &-                  &Insert/Delete          &\checkmark         &-                  &Low                    &High\\
\bottomrule
\end{tabular}%
}
\end{table*}

Table~\ref{tab:index_capabilities} summarizes popular spatial indexing libraries and implementations, in terms of the following features: the supported data types and number of dimensions, construction and maintenance operations, query operations, code complexity and extendibility.

\stitle{Boost Geometry} \cite{boost}. Geometry is a popular geospatial extension to the Boost library. Out of the common spatial data structures, the library includes only an R-tree implementation, offering however the entire range of features and capabilities. Specifically, the R-tree is an N-dimensional index supporting both point and MBB data. It offers bulk-loading construction and supports insertions, deletions for maintenance. For querying, both range and $k$NN search methods are available. On the other hand, the implementation is complex, requiring extensive knowledge of the Boost codebase. Specifically, the R-tree header alone contains over 2K lines and directly includes over 50 headers, many of them for internal components such as, node representation, visitors, updates, etc. Thus, mastering of the general Boost codebase is necessary for possible changes in the structure and functionality of the index. Hence, extending the R-tree structure or writing new search algorithms that use it is not straightforward since the implementation is tightly coupled with library-specific storage, shape, visitor, and query abstractions. Another inflexibility is that the developer cannot set a different capacity between the inner tree nodes and the leaves.
% \nikos{provide convincing details, e.g., thousands of files and dependencies} 

\stitle{CGAL} \cite{cgal:project}. The CGAL library includes implementations for Orth-trees
(CGAL's terminology for N-dimensional generalizations of Quad-trees and Oct-trees)
%\nikos{I do not know what an orth-tree is. Google gives me nothing. provide a citation.} 
and KD-trees as N-dimensional indices. Although only points are supported, the key drawback is that the library has limited support for dynamic operations and exact querying. Its Orth-tree is entirely static (supporting bulk-loading but no insertions or deletions), and its range queries return the intersected leaf nodes rather than the matching objects. Likewise, while the KD-tree implementation supports bulk-loading construction, range, and $k$NN queries, its insertion mechanism invalidates the structure, requiring a full rebuild before querying, rather than dynamically maintaining the tree. The KD-tree supports lazy deletions as re-balancing is not implemented. Finally, since CGAL is a comprehensive geometry library, it consists of a large interwoven codebase, making it hard to modify without deep knowledge of the library.

\stitle{GEOS} \cite{geos}. The GEOS library offers implementations for the R-tree and the Quad-tree to index both points and MBBs, but only in the 2D space. In particular, the R-tree implementation supports STR packing. 
After the tree is built, items may not be added or removed. Regarding querying, it supports range queries, but its nearest-neighbor search is limited to 1-NN queries. The library implements the MX-CIF Quad-tree variant that directly indexes MBBs and boxes\eat{spatial objects}, while points are represented as degenerate MBBs where the upper and lower bound at each dimension coincide. It supports dynamic insertion and deletion, but lacks native support for $k$NN queries and bulk-loading. 
Furthermore, its range query only prunes the search to nodes whose extent overlaps the query range, thus, the user must write custom post-filtering code to identify the exact query results.
Moreover, both implementations are tightly integrated into the GEOS geometry engine and rely on library-specific elements, thus increasing code complexity and making the indices difficult to extend. Finally, its memory overhead proved too high for large datasets, resulting in occasional out-of-memory crashes during our evaluation.

\stitle{LibSpatialIndex} \cite{libspatialindex}. The library provides a feature-complete R-tree implementation supporting bulk-loading construction, insertions and deletions for maintenance, 
%\nikos{does it have different options for insertion/deletion, for example R*-tree insertions, quadratic split, etc.? We need to be specific of the update algorithms that it supports}, 
range and $k$NN queries over N-dimensional objects (points and MBBs). 
Specifically for dynamic updates, it supports the linear, quadratic, and R*-tree splitting. Although the index can reside in main-memory, the code architecture is disk-oriented which introduces substantial overhead, making it significantly slower than the in-memory alternatives. Moreover, while its codebase is less complex than Boost's, it is not straightforward to adapt since the R-tree implementation is similar to Boost's tightly coupled with library-specific storage, shape, visitor, and query abstractions.

\stitle{PCL} \cite{pcl}. The Point Cloud Library (PCL) provides Oct-tree and KD-tree implementations for 3D points. Both store coordinates as float values, making direct comparison against indices that support double-precision coordinates hard. For range queries in the Oct-tree, this limitation is addressed by expanding each query bound by one representable float value and post-filtering the returned candidates against the original double-precision coordinates. However, this single-precision representation may still misrank candidates for $k$NN queries, where post-filtering is not sufficient to guarantee the exact double-precision result. 
%Consequently, we exclude PCL $k$NN from our experiments. 
PCL's KD-tree is based on FLANN \cite{DBLP:journals/pami/MujaL14}. It supports bulk-loading construction but does not expose insertion or deletion through the PCL wrapper. In contrast, PCL's Oct-tree is populated through point insertions rather than a packed bulk-loading procedure, and its deletion API removes the entire leaf addressed by a point, instead of the individual point record. %\nikos{what does fine-grained point deletion mean? be specific instead of using fancy keywords}.

\stitle{Nanoflann} \cite{nanoflann}. Nanoflann is a popular KD-tree implementation supporting points in an $N$-dimensional space. It provides distinct index adaptors for static and dynamic data. The static tree supports bulk-loading construction and exact window range queries. 
In contrast, the dynamic adaptor (which is implemented as a forest of static KD-trees) supports incremental insertions and deletions, but only exposes a radius-shape range query interface. 
Furthermore, its construction is handled by grouping points into bins that are individually bulk-loaded into multiple KD-trees, while point deletions are handled lazily using tombstones.
A workaround to evaluate window range queries on the dynamic adaptor is to perform an enclosing-ball radius search and then post-filter the returned candidates against the exact query MBB. 

\stitle{Vigier} \cite{pvigier}. Vigier's library implements an MX-CIF Quad-tree, supporting both points and MBB data. Crucially, it does not support bulk-loading construction and requires an a priori global bounding box upon initialization, as it cannot dynamically expand to accommodate out-of-bounds insertions. Furthermore, it supports spatial range queries but lacks $k$NN search.

\section{The Indexicon Library}
\label{sec:lib}
We now describe our Indexicon library for spatial indexing which directly tackles the challenges listed in Section~\ref{sec:intro}. Indexicon is open-sourced at \href{https://github.com/psimatis/Indexicon-Spatial-Library}{https://github.com/psimatis/Indexicon-Spatial-Library}.

\subsection{Code architecture and API}
A primary goal of Indexicon is to avoid the fragmented interfaces prevalent in existing libraries. Thus, all indices adhere to a common architectural pattern and a unified C++ interface, regardless of their underlying algorithms. As illustrated in Listing~\ref{lst:indexicon_api}, each index is instantiated as a standalone C++ template, parameterized by the coordinate type, payload identifier type, dimensionality, and node capacity. This avoids the heavy, interwoven class hierarchies found in libraries like Boost or CGAL, allowing developers to drop the headers directly into their projects. The interface standardizes all core spatial operations. Indices can be dynamically populated from an empty state or efficiently bulk-loaded using an iterator range. Dynamic maintenance is handled via \texttt{insert} and \texttt{remove} methods. For data retrieval, the API supports both window range queries (\texttt{rangeQuery}) and $k$-nearest neighbor searches (\texttt{knnQuery}). Search results are directly emitted to a provided C++ result container. This allows developers to collect results in any standard container (e.g., \texttt{std::vector}), avoiding the overhead of proprietary result objects. Finally, a \texttt{getStatistics} function exposes internal structural metrics, including tree height, node counts, and memory footprint, facilitating transparent benchmarking.

\begin{lstlisting}[
    language=C++, 
    caption={The API template of Indexicon}, 
    captionpos=b, 
    label={lst:indexicon_api}, 
    basicstyle=\ttfamily\scriptsize, 
    breaklines=true, 
    float=t,
    % 1. List your function names here:
    emph={SpatialIndex, insert, remove, rangeQuery, knnQuery, getStatistics},
    % 2. Style them magenta:
    emphstyle=\color{magenta}
]
template <typename Coord, typename Value, unsigned int Dims, unsigned int Cap>
class SpatialIndex {
public:
    SpatialIndex();
    
    // Bulk-loading constructor
    template <typename Iter>
    SpatialIndex(const Iter first, const Iter last);

    // Maintenance
    void insert(const Point p); 
    void insert(const MBB b); 
    bool remove(const Point p);
    bool remove(const MBB b); 

    // Range search
    template <typename Result>
    Result rangeQuery(const MBB q);

    // kNN Search
    template <typename Result>
    Result knnQuery(const Point q, const unsigned int k);
    
    // Structural metrics
    Stats getStatistics();
};
\end{lstlisting}

\eat{
\nikos{does this paragraph say the same things as the previous one? I see quite a lot of overlap. I suggest merging them to a single paragraph}\panos{yes it does, I wanted to remove it}
The current landscape of spatial indexing libraries is highly fragmented. Researchers frequently encounter implementations burdened with complex multi file dependencies or those that critically lack essential operations. To address this, Indexicon is designed to be highly portable and effortless to integrate. All indexes in our library share a common architecture guided by strict design rules. Each index is a self contained, templated C++ header only implementation with absolutely no external dependencies beyond the standard library. The coordinate type, dimensionality, and node capacity are all compile time parameters. To ensure researchers have a complete toolkit, every index uniformly supports bulk-loading, dynamic insertion and deletion, range queries, $k$ nearest neighbor ($k$NN) search, and structural statistics reporting. 
Table~\ref{tab:index_capabilities} summarizes the capabilities of each spatial index in Indexicon, alongside alternative open-source implementations.
}

All indexing structures of Indexicon are designed to reside entirely within main memory. This design focuses on ultra-low latency in-memory traversal over disk I/O, recognizing that I/O is no longer the primary performance bottleneck in modern commodity hardware~\cite{DBLP:journals/pvldb/LarsonL16, DBLP:conf/sigmod/MagalhaesBM23, amagata2026firas, DBLP:conf/sigmod/0005BM22}. Similar to other existing spatial libraries~\cite{geos, boost, cgal:project, pcl, libspatialindex}, our implementation operates on a single-threaded execution model. 
This provides sufficient throughput for typical workloads and yields lean, highly-readable search performance for typical spatial search tasks.
These deterministic execution paths currently serve as excellent raw performance baselines and embeddable components for spatial execution engines. 
Extending our library to fully support multi-threaded execution is a primary objective for our future work. 

\subsection{Spatial indexing}
\label{sec:indexes}
The current implementation of Indexicon includes the most popular tree-based spatial indices for low-dimensional points and minimum bounding boxes (MBBs), which support the filter step of spatial queries. We describe these data structures and how they are implemented within Indexicon.

%\subsubsection{The R-tree}
%The \textbf{R-tree}~\cite{DBLP:conf/sigmod/Guttman84} 
\stitle{R-tree}. Our implementation
is an N-dimensional index supporting both point and MBB data. It natively supports decoupled maximum capacities (fanouts) for internal nodes and leaf nodes, enabling developers to independently optimize cache alignment and structural depth. Its bulk-loading mechanism mirrors the packing strategy \cite{DBLP:conf/gis/GarciaLL98} adopted by the Boost Geometry library \cite{boost}. Instead of relying on rigid bottom-up sort-tile-recursive scheme \cite{DBLP:conf/icde/LeuteneggerEL97}, it constructs the tree top-down utilizing a recursive KD-tree-like bisection. At each step, it identifies the longest spatial axis of the bounding box and employs a median selection algorithm to evenly divide the records. Insertions are implemented based on the R*-tree methodology \cite{DBLP:conf/sigmod/BeckmannKSS90} featuring a forced re-insertion mechanism. When a node overflows, rather than splitting it immediately, the index temporarily removes a fraction (30\% by default) of its entries farthest from the node center and re-inserts them from the root to continuously refine the overall tree structure. If a node must split, the index comprehensively evaluates all dimensions, computes the margin sums for every possible partition, and selects the axis that minimizes the total margin. Finally, deletions handle structural underflow by dissolving sparse nodes, gathering their remaining entries as orphans, and recursively reinserting them. 

%\subsubsection{Quad-tree variants}\label{sec:quadtree}
%Our \textbf{Quad-tree} 
\stitle{Quad-tree \& variants}. Our implementation of the Quad-tree
is a 2D structure for point data with bucket leaves. We developed it with three possible splitting strategies. The most popular strategy, followed by the \emph{point region} (PR) Quad-tree~\cite{DBLP:journals/csur/Samet84},
divides each overflowing node at the geometric midpoint of its bounding box, producing four spatially equal quadrants. The \emph{pseudo-median} strategy~\cite{DBLP:journals/acta/OvermarsL82} splits using the independent medians of the x and y axes. Finally, the \emph{point longest-axis} strategy splits along the median of the longest span while using the geometric midpoint for the other axis. Bulk-loading constructs the tree top-down using a divide-and-conquer approach. Starting with the global bounding box and the full set of points, the algorithm computes a split point according to the chosen strategy and creates four child nodes. The dataset is then partitioned into four and 
% subsets based on spatial intersection. This 
the process is recursively applied to each partition and child node, terminating only when the number of objects in the partition does not exceed the leaf node capacity threshold, 
% a subset size falls within the configured node capacity, 
at which point the partition becomes a leaf. 
% If the data are MBBs, they are replicated to all partitions they overlap with. \rapper{not true. the quad tree doesnt support MBBs. only the mxcif one does}
All Quad-tree variants handle extreme data skew by allowing overflowing leaves when numerous data objects have identical coordinates, instead of recursing infinitely. An optional maximum depth limit can be imposed, if needed for an application, though our experiments show that the tree naturally stops subdividing once buckets fall below capacity, making an artificial limit unnecessary. 

To support data in a dynamic spatial domain, where minimum and maximum coordinates in each dimension are not fixed, dynamic insertions expand the tree via an automatic re-rooting mechanism. Whenever a new point falls outside the spatial domain currently covered by the tree, the index generates a new root that expands the spatial domain in the direction of the outlier. The entire existing tree is demoted to become one of the four child quadrants of this new root, while the remaining three siblings are initialized as empty nodes. This constant-time topological shift repeats iteratively until the bounding box of the new root fully encompasses the incoming point. Notably, while median-based strategies guarantee structural balance during bulk-loading, the PR strategy and dynamic insertions lack re-balancing mechanisms, meaning clustered data can produce long search paths. Deletion merges sibling leaves back into their parent when their combined count falls below capacity.

Our Oct-tree\eat{~\cite{DBLP:journals/cvgip/Meagher82}} extends the PR Quad-tree decomposition to three dimensions, splitting space into eight equal octants at the midpoint of each axis. 
%\nikos{why not use medians also in the 3D space? you need to justify this rigidness of the Oct-tree} \rapper{because the PR split performed well according to the experiments}
%\nikos{still, if you want to support this as a functionality, we have to put it in the future work}
It shares the same bucket leaf design and identical coordinate duplicate handling safeguard as the 2D Quad-tree. Furthermore, it seamlessly mirrors the dynamic capabilities of the 2D structures by fully supporting re-rooting insertion for handling out of bounds points and consolidating sibling leaves upon deletion.

Our MX-CIF Quad-tree\eat{~\cite{DBLP:conf/dac/Kedem82}} is a 2D structure specialized for MBB data, which avoids data replication.
%\nikos{why not for 3D as well? At the end of the paper, include a list of future work items for things that have not been implemented but should be part of a future version of Indexicon library.} \rapper{didnt think about it}
%\nikos{add to the future work in the end}
Unlike the point Quad-tree, MBBs that straddle a split boundary are stored at the internal node itself rather than being pushed to a child. Straddling MBBs are maintained in sorted order along their secondary axis during bulk-loading. Non-straddling MBBs are placed in bucket leaves, a design choice that significantly improved range query performance over the classical unbucketed layout. Bulk-loading classifies each MBB as straddling the vertical midline, straddling only the horizontal midline, or fitting entirely within one quadrant. The first two groups are stored at the current node while the quadrant groups are recursed into child nodes. Dynamic insertion also supports re-rooting for expanding spatial domains.

%\subsubsection{The KD-tree}
%The \textbf{KD-tree}~\cite{DBLP:journals/cacm/Bentley75} 
\stitle{KD-tree}. Our implementation
is an N-dimensional spatial index designed for point data, featuring a bucket leaf architecture. It is implemented as a binary space partitioning tree where each internal node splits the spatial domain along a single axis-aligned hyperplane. During bulk-loading, the index rapidly builds a balanced structure by recursively selecting a splitting dimension and 
{\em cracking} \cite{DBLP:conf/cidr/IdreosKM07}
% cleanly dividing \nikos{what is ``cleanly dividing''? Do you perform cracking/swapping using the median?} \rapper{what is cracking and swapping? i use std::nth\_element which rearranges the records in place.}
% \nikos{this is probably the same as cracking, i.e., one step of quicksort (bouros, can you please confirm?)}
the dataset at the median coordinate of the chosen split axis. This recursive in-place partitioning continues until the number of points falls below the predefined node capacity, at which point a bucket leaf is instantiated to store those records. The dimension selection process is controlled by a compile-time configuration that offers three distinct strategies. By default, the index uses an adaptive approach that dynamically calculates the spatial spread of the data points across all dimensions and splits along the axis exhibiting the widest spatial extent \cite{DBLP:journals/toms/FriedmanBF77}, actively mitigating the structural impact of skewed data distributions. Alternatively, this can be toggled to a round-robin strategy which cycles through the dimensions based on the current tree depth, or a longest-axis strategy that simply splits the widest dimension of the node's geometric bounding box. Dynamic insertions traverse the established splitting hyperplanes to map the incoming point directly to the appropriate bucket leaf. If an insertion causes the bucket to exceed its maximum capacity, the leaf immediately splits along a selected dimension, extending the binary hierarchy downward. It should be noted that, unlike bulk-loading, sequential insertions do not perform global rebalancing; consequently, highly skewed insertion workloads may eventually degrade the tree's structural balance. 
%\nikos{another point to be included in the future work. implement rebalancing for the KD-tree. this can be done by reconstructing subtrees that cause imbalancing; you may talk to Achilleas who has implemented this.}
Finally, deletions operate by executing a standard point lookup and performing a linear scan within the matching bucket leaf to securely erase the targeted record. While the index avoids the computational overhead of global tree restructuring, it actively mitigates underflow. Specifically, if two sibling leaves experience enough deletions that their combined element count falls below the node capacity, they are merged back into their parent, converting it back into a single bucket leaf.
% \nikos{this may also cause imbalance. again, a point for future work}

\begin{figure*}[t]
    \centering
    \includegraphics[width=\textwidth]{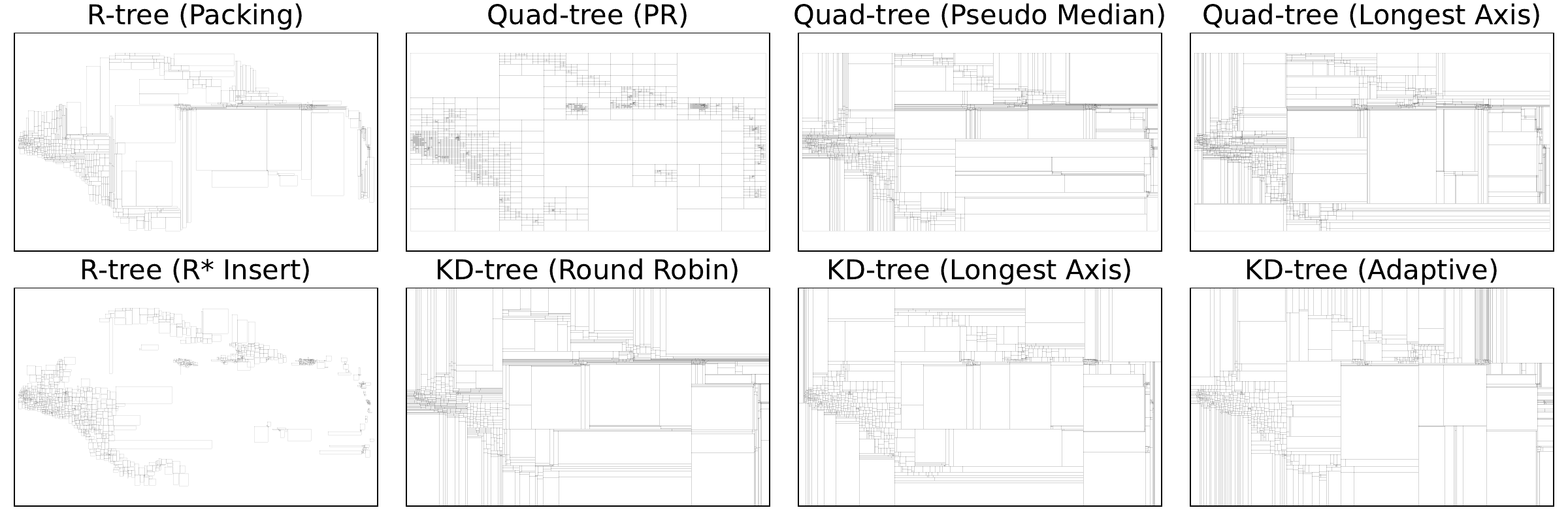}
    \caption{The leaf MBBs generated across our various indexing strategies, executed on a 100K point sample of OSM dataset.}
    \label{fig:partitioning_strategies}
\end{figure*}

% \subsubsection{Visualization of leaf node boundaries}
\stitle{Visualization}.
Figure~\ref{fig:partitioning_strategies} visualizes the leaf-level spatial partitions generated by Indexicon's indices, executed on a 100K-point randomly drawn sample of the OSM dataset with a node capacity of 128. For the R-tree, the top-down packing strategy yields tightly grouped, minimally overlapping bounding boxes resulting in a highly compact structure of 782 leaves. In contrast, dynamic R*-tree insertions produce more irregular, overlapping bounds with minimal margins, expanding the tree's footprint to 1,140 leaves. 
%\panos{Should we reference here previous visualizations and say that we make similar observations? Papadias had a similar figure in one of his papers I believe} 
The Quad-tree variants illustrate the difference between the strict, geometrically uniform quadrants of the Point-Region (PR) strategy (which generates the most fragmented space at 2,098 leaves) and the tighter, data-driven bounds of the Pseudo Median (1,654 leaves) and Longest Axis (1,996 leaves) techniques. Finally, the KD-tree visualizations highlight the structural mechanics of binary space partitioning. Because they strictly halve the data, all three KD-tree variants generate 1,024 leaves. However, the visualization demonstrates how the Adaptive and Longest Axis strategies conform to the spatial spread of the data better than the alternating splits of the Round Robin approach.
%(e.g., notice the long-thin boxes over Mexico).

\subsection{Spatial query processing}

All indices share the same range query algorithm: a recursive descent that, at each internal node, prunes children whose bounding region%
\footnote{For Quad-trees and KD-trees each node stores an MBB for all data in its subtree.}
does not intersect the query. When the query's MBB fully encloses a node's bounding region, all data items in the subtree are scanned and reported without comparisons, thus eliminating redundant computations in large-area queries. In the MX-CIF Quad-tree, for each visited internal node all data items in its straddle list are compared to the query range. 
%The Grid index achieves the same effect iteratively: it identifies the range of cells overlapping the query window and applies the containment shortcut per cell.

The $k$-nearest neighbor ($k$NN) search in all indices follows a best-first traversal. The algorithm maintains a priority queue of unvisited nodes ordered by their minimum squared distance to the query point, together with a max-heap containing the current $k$ best candidate neighbors. At each step, the nearest unvisited node is removed from the node queue. If its minimum distance is no smaller than the distance of the current $k$-th nearest candidate, the search terminates, since all remaining queued nodes are at least as far. Otherwise, leaf entries are scanned and the candidate heap is updated. For internal nodes, children whose minimum distance can improve the current result are inserted into the node queue. In the MX-CIF Quad-tree, visited internal nodes additionally scan their straddle lists during traversal.
%\panos{If this is a best-first strategy (Hanan's) you don't need the max-heap; you stop after popping $k$ data elements from the min-heap} 
% \nikos{This is not accurate. I suggest rephrasing to ``as soon as the top node in the min-heap is further from the query point compared to the current $k$-th NN (i.e., the top element of the max-heap), search terminates.''}

% The Grid index uses a similar best first strategy adapted for flat cells: it utilizes a priority queue to explore cells by minimum distance and strictly requires a hash set to track visited cells, terminating once the minimum distance to the next unvisited cell exceeds the farthest current neighbor.
% \nikos{for grid, the best strategy includes cell-groups (in the same rows or  column) in the heap in addition to single cells. See mouratidis paper (conceptual partitioning)}

%% --- SECTION 3 ---

\section{Comparative Study}\label{sec:study}

\begin{figure*}[t]
    \centering
    \includegraphics[width=\textwidth]{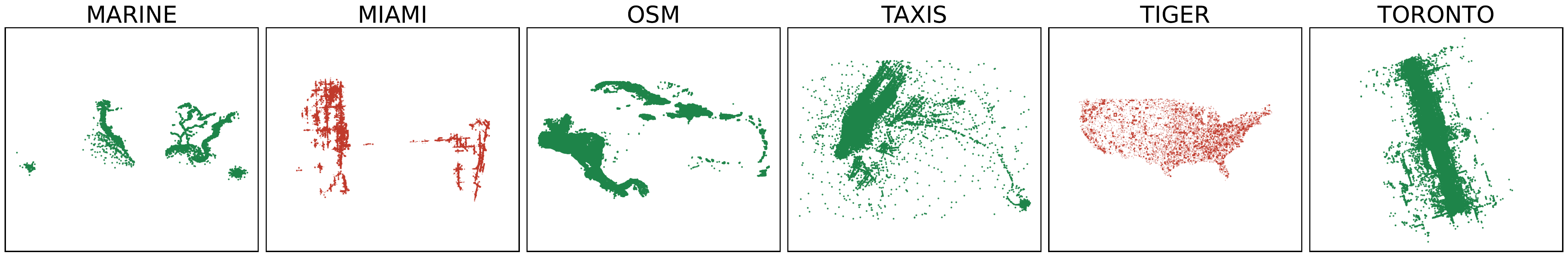}
    \caption{Spatial distribution of the datasets used in our evaluation  projected onto their first two dimensions, with point and MBBs of geometries plotted in green and red, respectively}
    \label{fig:dataset_scatter}
\end{figure*}

In this section, we experimentally compare the spatial indices in our Indexicon library with their most popular and best-performing open-source implementations. We first present the datasets and queries in our tests, then describe the competitors, and finally present the experimental results. All experiments were conducted on a workstation running Ubuntu 24.04.3 LTS, equipped with an Intel Core i7-14700K processor (33 MB L3 cache) and 128 GB of main memory. All codes are in C++ and compiled using GCC version 13.3.0 with the -O3 optimization flag. All datasets, spatial indices, and query workloads reside entirely in main memory. 

\subsection{Experimental setup}

\begin{table}
\centering
\caption{Dataset statistics and structural profiles}
\label{tab:datasets}
\resizebox{\columnwidth}{!}{%
\begin{tabular}{lrlrcrp{4.5cm}}
\toprule
\textbf{Dataset} & \textbf{Records} & \textbf{Type} & \textbf{Size} & \textbf{Dims} & \textbf{Dupl.} & \textbf{Description} \\
\midrule
MARINE & 25.0M & Point & 716.2 MB & 3D & 0.01\% & US coastal vessel tracking data \\
MIAMI & 3.5M & MBB$\to$Point & 312.2 MB & 3D$\to$2D & 0.02\% & Urban traffic-object MBBs in Miami \\
OSM & 103.5M & Point & 2.0 GB & 2D & 0.03\% & Geolocations in Central America \\
TAXIS & 112.8M & Point & 2.2 GB & 2D & 14.55\% & NYC Taxi pickup geolocations \\
TIGER & 17.9M & MBB$\to$Point & 715.2 MB & 2D & 5.60\% & Lower 48 street MBBs \\
TORONTO & 21.6M & Point & 679.5 MB & 3D$\to$2D & 6.94\% & Toronto urban LiDAR point cloud \\
\bottomrule
\end{tabular}%
}
\end{table}

\stitle{Datasets.} We use six real-world datasets spanning various geographic applications and data types; Table~\ref{tab:datasets} lists their \eat{dataset }statistics and Figure~\ref{fig:dataset_scatter} visualizes them. Tailored for spatial and spatiotemporal applications, our evaluation focuses on 2D and 3D \eat{geographic }domains.

\begin{itemize}
% [noitemsep,topsep=2pt,parsep=0pt,partopsep=0pt,leftmargin=0.5cm]
    \item \textbf{MARINE}~\cite{marine}:
    Contains Automatic Identification System (AIS) spatiotemporal (i.e., 3D) tracking data for vessels in US coastal waters, characterized by trajectories clustered along shipping lanes.
    
    \item \textbf{MIAMI}~\cite{Argoverse2}: Contains 3D MBBs of annotated traffic objects in Miami, such as a vehicle, cyclist, or pedestrian.
    
    \item \textbf{OSM}~\cite{OpenStreetMap}: Contains 2D spatial points in Central America from OpenStreetMap.
    
    \item \textbf{TAXIS}~\cite{taxis}:
    Captures 2D spatial points representing NYC taxi pickup locations, exhibiting hotspots in dense urban zones.
    
    \item \textbf{TIGER}~\cite{tiger}:
    Consists of 2D MBBs tracking local street centerlines across the contiguous United States.
    
    \item \textbf{TORONTO}~\cite{tan2020toronto3d}: A dense and highly clustered 3D LiDAR point cloud dataset capturing urban roadways in Toronto, Canada.
\end{itemize}
% \noindent To conduct standardized 2D benchmarking across the entire collection, we project the higher-dimensional (MARINE and TORONTO) and the MBB datasets (MIAMI and TIGER) onto a 2D plane by utilizing their first two spatial dimensions. This uniform mapping allows us to comprehensively evaluate standard 2D spatial indexing structures across contrasting geometric primitives. 

\noindent To conduct standardized 2D point benchmarking across the entire collection, we derive a 2D point representation from each dataset by retaining only its first two coordinates. For originally 3D point datasets, this corresponds to projecting each point onto its first two dimensions. For MBB datasets, we use the first two coordinate attributes to obtain representative 2D points. This gives us a uniform 2D point workload while preserving the original data sources and spatial distributions.

\stitle{Benchmark workload}.
Unless stated otherwise, we evaluate bulk-loading and maintenance in a single mixed workload. We randomly bulk-load 80\% of each dataset, insert the remaining 20\% sequentially, and then delete 5\% of the records. This models a common setting where an index is first built over historical data and later maintained under incoming updates, with deletions occurring less frequently. For methods without bulk-loading support (Table~\ref{tab:index_capabilities}), we build the initial 80\% by one-by-one insertions and report the cumulative insertion time as the bulk-loading cost.

\stitle{Queries.}
We generate 1,000 queries for each query type, including range, and $k$-nearest neighbor ($k$NN) queries. For this, we randomly sample points 
without replacement directly from the dataset, ensuring the query workload follows the underlying data distribution. For $k$NN queries, these are used as query points, and we vary $k$ in  $\{1, 10, 100\}$. Range queries are formulated as axis-aligned hyper-rectangles with side lengths proportional to the domain extent of each respective dimension, using as geometric centers a fresh random sample of dataset points. We configure their extents to cover 0.01\%, 0.1\%, and 1.0\% of the domain at each axis.
%, and we center them on a fresh random sample of dataset points drawn with replacement. 
%Furthermore, to evaluate performance under an out-of-distribution workload, we generate an additional set of queries where the points (or centers for range queries) are drawn uniformly at random from the domain space, independent of the underlying data distribution. 
%\nikos{do you include results for out-of-distribution queries? if not, delete the last sentence.}
Finally, we report the total execution latency per query setting.

\stitle{Competitors}.
%Selecting competitors is not straightforward, because the libraries expose different functionalities or target different data types, dimensionalities, or query semantics (see Table~\ref{tab:index_capabilities}). We therefore compare a
We select the competitors according to their ability to handle our workload. Therefore, we compare Indexicon
against the closest usable implementation for each index family\eat{ for a workload that requires insertion capabilities}. For the R-tree, we consider Boost Geometry, since it supports the full mixed workload: bulk-loading, insertions, deletions, range queries, and $k$NN search. We do not consider GEOS's STR-tree because it is immutable after construction and supports only 1-NN search. Furthermore, we do not consider LibSpatialIndex because its disk-oriented architecture makes it less representative of the lightweight main-memory setting evaluated here; a previous evaluation in \cite{LoskotW19} also demonstrated that Boost R-tree outperforms LibSpatialIndex R-tree. For point Quad-trees and MX-CIF Quad-trees over MBBs, we consider GEOS, applying the required post-filtering step to obtain exact range-query results. For 3D point indexing, we consider PCL's Oct-tree. However, due to its single-precision representation, we only consider PCL for range-search experiments and exclude it from $k$NN comparisons. For KD-trees, we consider Nanoflann, evaluating its dynamic adaptor since we assume non-static workloads. We exclude CGAL because its indices lack on maintenance, and its Orth-tree reports intersected leaves rather than exact objects. We also exclude Vigier's Quad-tree because it requires a fixed global domain at initialization, and does not support $k$NN search.

\subsection{R-tree benchmark}

%In this section, we compare Indexicon's R-tree implementation with the Boost R-tree. 
In the first set of experiments, we investigate how parameter choices (node capacity, percentage of removed entries at forced reinsert) affect the performance of the two R-tree implementations. Then, we compare Indexicon and Boost R-tree implementations in construction and update costs as well as in query performance.

\begin{figure*}
    \centering
    \begin{subfigure}{\textwidth}
        \centering
        \includegraphics[width=\textwidth]{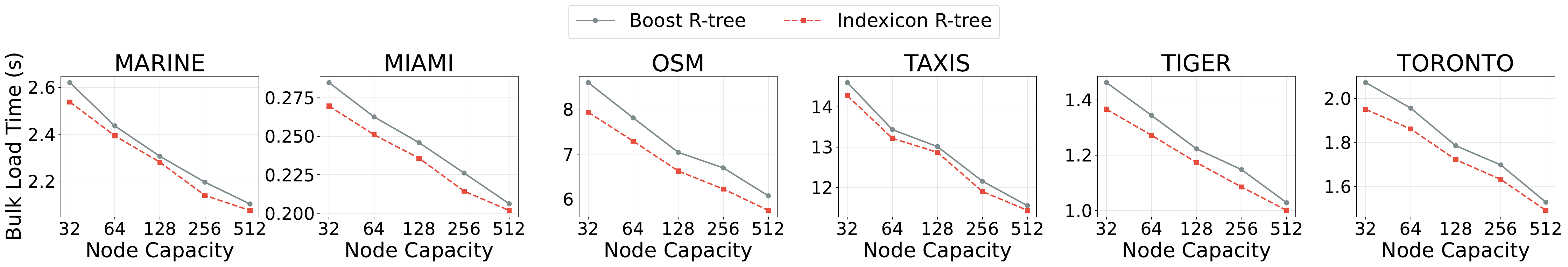}
        \caption{Bulk-loading}
        \label{fig:capacity_bulk}
    \end{subfigure}
    
    \vspace{1.5em}
    
    \begin{subfigure}{\textwidth}
        \centering
        \includegraphics[width=\textwidth, trim=0pt 0pt 0pt 75pt, clip]{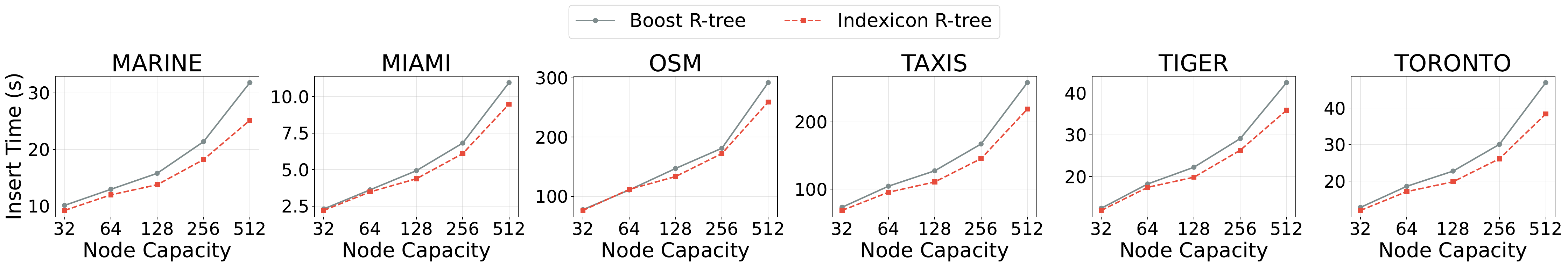}
        \caption{Dynamic insertions}
        \label{fig:capacity_insert}
    \end{subfigure}
    
    \vspace{1.5em} 
    
    \begin{subfigure}{\textwidth}
        \centering
        \includegraphics[width=\textwidth, trim=0pt 0pt 0pt 75pt, clip]{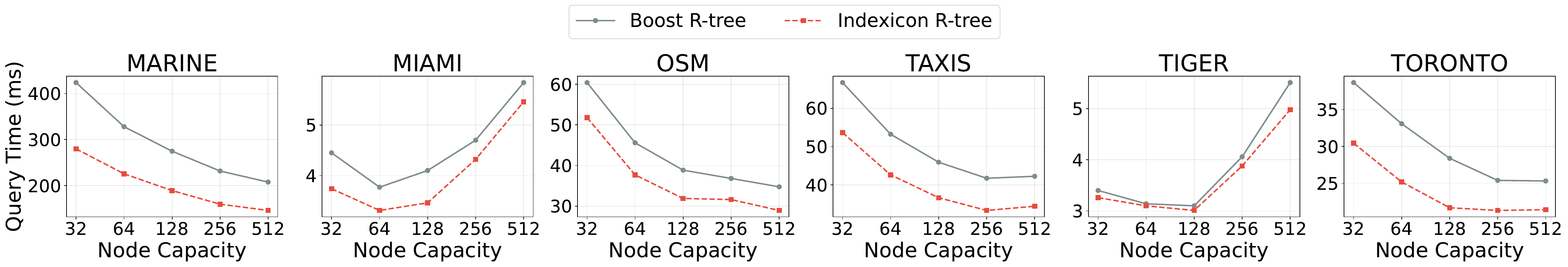}
        \caption{Range queries of 0.1\% extent}
        \label{fig:capacity_query_0_10pct}
    \end{subfigure}
    
    \caption{Impact of R-tree node capacity on bulk-loading, dynamic insertions, and range query performance; workload 80\% of each dataset bulk-loaded, 20\% insertions and 1000 queries}
    \label{fig:capacity_impact}
\end{figure*}

\subsubsection{Effect of parameters}
% We now investigate the mechanics of the R-tree to optimize its performance. Specifically, we evaluate critical aspects of node capacity, and the impact of the R*-tree forced reinsertions during overflows.
An important parameter in R-tree configuration is the maximum node capacity (or fanout), which dictates the maximum number of entries a node can hold. While Indexicon natively supports setting distinct capacities for internal nodes and leaves, allowing developers to optimize for different cache-line alignments, Boost's R-tree enforces a single capacity for both. To ensure a fair comparison, we constrain Indexicon to use identical capacities for both node types in this experiment. We ran this experiment with a workload of 80\% bulk-loading and 20\% insertions.

Figure~\ref{fig:capacity_impact} visualizes the performance impact of varying the node capacity from 32 to 512 across our datasets. The results highlight an inherent trade-off between construction speed, insertion scalability, and query efficiency for both implementations. Importantly, Indexicon consistently outperforms Boost across all configurations. Figure~\ref{fig:capacity_bulk} shows that increasing the node capacity accelerates bulk-loading. A higher capacity yields a shallower tree with fewer nodes in total, significantly reducing memory allocation overhead and recursions during packing. Conversely, dynamic insertion (Figure~\ref{fig:capacity_insert}) scales poorly with larger capacities. High-capacity nodes incur a heavier penalty during inserts, as the algorithms must evaluate larger arrays to select the optimal subtree, and perform more extensive memory shifts during node overflows. Finally, Figure~\ref{fig:capacity_query_0_10pct} demonstrates that spatial search benefits from medium node sizes. Larger capacities consolidate data into fewer, contiguous blocks of memory, minimizing tree depth and cache misses during traversal. Across all datasets, the 0.1\% range query latencies drop significantly as capacity increases, though the performance gains begin to degrade or plateau beyond a capacity of 128 for multiple datasets (e.g., TIGER). This degradation occurs because large nodes force the query to linearly scan irrelevant entries within the leaves, offsetting the benefits of a shallower tree. Thus, we select capacity 128 as the default value for the rest of our R-tree benchmark.

\begin{figure*}
    \centering
    \begin{subfigure}{\textwidth}
        \centering
        \includegraphics[width=\textwidth]{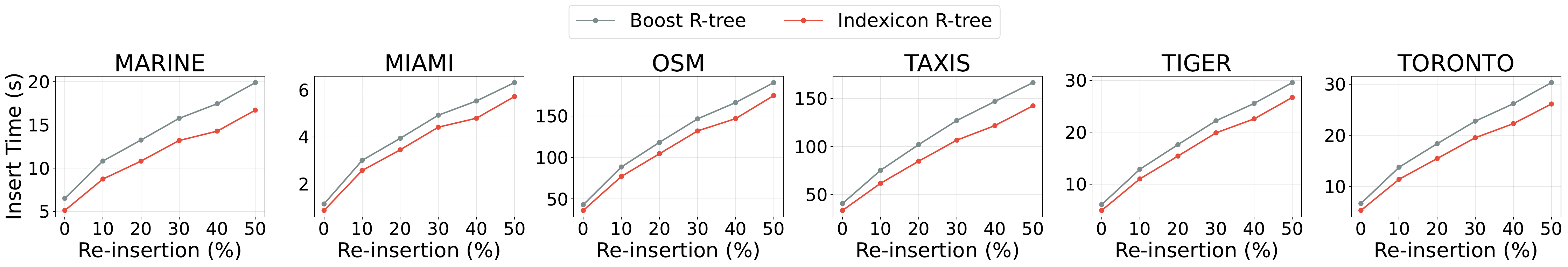}
        \caption{Dynamic insertions}
        \label{fig:reinsertion_insert}
    \end{subfigure}
    
    \vspace{1em} 
    
    \begin{subfigure}{\textwidth}
        \centering
        \includegraphics[width=\textwidth, trim=0pt 0pt 0pt 75pt, clip]{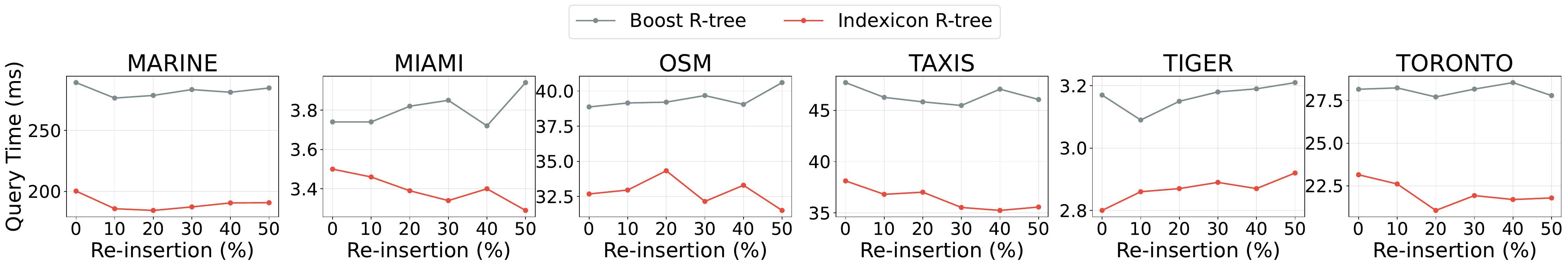}
        \caption{Range queries of 0.1\% extent}
        \label{fig:reinsertion_query}
    \end{subfigure}
    \caption{Impact of R-tree forced re-insertion percentage on 2D point datasets\eat{\panos{point, MBBs or both}}; workload 80\% of each dataset bulk-loaded, 20\% insertions and 1000 queries}    
    \label{fig:reinsertion_impact}
\end{figure*}

Next, we explore the efficacy of forced re-insertion on 2D point data. In the R*-tree, when a node overflows, the index temporarily removes a predefined fraction of elements farthest from the node center and re-inserts them from the root to refine the tree structure. Like in the capacity experiment, the selected workload consists of 80\% bulk-loading and 20\% insertions\footnote{We repeated the experiment with a fifty-fifty split between bulk-loading and insertions and observed similar results.}. Figure~\ref{fig:reinsertion_impact} illustrates the impact of varying the re-insertion percentage from 0\% up to 50\% on dynamic insertion and 0.1\% range query times. These behavioral trends were verified across both our Indexicon and Boost, with our implementation consistently maintaining superior performance across all configurations. The results reveal that increasing the re-insertion percentage imposes a severe penalty on insertion latency. On OSM, disabling re-insertion (i.e., 0\%) yields an insertion time of 36.04 secs, whereas utilizing the historically accepted 30\% nearly quadruples the time to 131.95 secs.

Conversely, the anticipated improvements in query execution times are marginal and highly inconsistent. As observed in Figure~\ref{fig:reinsertion_query}, the query times exhibit slight fluctuations rather than demonstrating a smooth, monotonic improvement as the re-insertion rate increases. This disconnect stems from the architectural shift from disk-bound to main-memory systems. The 30\% heuristic was originally optimized for magnetic disks, where spending CPU time during insertions to improve tree structure was justified if it saved disk reads during a query. However, for an in-memory framework, burning CPU cycles to re-insert entries causes huge insertion delays, while the structural improvements result in negligible time-savings of memory traversal. The most significant drop in query time occurs on MARINE where it falls from 200.35 msecs (at 0\%) to 185.94 msecs at a 10\% reinsertion rate, but pushing the rate higher yields no benefit. 
% \panos{what drops? the ``it drops'' is unclear} 
Meanwhile, on datasets like TIGER, query performance remains effectively static (fluctuating trivially between 2.80 msecs and 2.89 msecs). Given the extreme computational cost incurred during insertions and the unpredictable, trivial returns observed during search operations, we conclude that the historically empirical standard of 30\% forced re-insertion is highly detrimental for modern workloads. Thus, we advocate for keeping re-insertion either strictly disabled (0\%) or limited to small fractions (e.g., 10\%, \eat{which we use} as the default) for the remaining experiments.

\subsubsection{Comparing R-tree implementations}
We now compare the performance of Indexicon's R-tree against the Boost Geometry R-tree across all datasets. The evaluation is conducted using the default mixed workload. We break down the analysis by data type and dimensionality.

\begin{table}[t]
\centering
\footnotesize
\setlength{\tabcolsep}{3pt}
\caption{R-tree packing, dynamic insertion and deletion times; default workload}
\label{tab:rtree_build_insert}
\begin{tabular}{lrrrrrr}
\toprule
& \multicolumn{3}{c}{\textbf{Boost R-tree}} & \multicolumn{3}{c}{\textbf{Indexicon R-tree}} \\
\cmidrule(lr){2-4} \cmidrule(lr){5-7}
\textbf{Dataset} & \textbf{Build (s)} & \textbf{Insert (s)} & \textbf{Delete (s)} & \textbf{Build (s)} & \textbf{Insert (s)} & \textbf{Delete (s)} \\
\midrule
\multicolumn{7}{l}{\textbf{2D point}} \\
\midrule
MARINE & 2.34 & 11.20 & 3.22 & 2.34 & 9.05 & 1.76 \\
MIAMI & 0.24 & 3.08 & 0.21 & 0.23 & 2.51 & 0.17 \\
OSM & 7.14 & 91.10 & 9.01 & 6.78 & 77.16 & 7.59 \\
TAXIS & 13.22 & 77.16 & 9.81 & 13.18 & 62.40 & 8.62 \\
TIGER & 1.20 & 13.30 & 1.08 & 1.16 & 11.24 & 0.94 \\
TORONTO & 1.78 & 14.33 & 1.51 & 1.72 & 11.53 & 1.21 \\
\midrule
\multicolumn{7}{l}{\textbf{3D point}} \\
\midrule
MARINE & 2.26 & 20.30 & 1.91 & 2.22 & 17.75 & 1.73 \\
MIAMI & 0.30 & 3.75 & 0.24 & 0.25 & 3.24 & 0.21 \\
TORONTO & 2.06 & 19.52 & 2.43 & 1.88 & 17.72 & 1.79 \\
\midrule
\multicolumn{7}{l}{\textbf{2D MBB}} \\
\midrule
MIAMI & 0.26 & 2.71 & 0.35 & 0.26 & 2.00 & 0.35 \\
TIGER & 1.25 & 10.88 & 2.18 & 1.24 & 7.71 & 1.91 \\
\midrule
\multicolumn{7}{l}{\textbf{3D MBB}} \\
\midrule
MIAMI & 0.28 & 4.66 & 0.40 & 0.28 & 2.52 & 0.39 \\
\bottomrule
\end{tabular}
\end{table}

Table~\ref{tab:rtree_build_insert} details the construction and maintenance latencies per object type and dimensionality.
Across the board, Indexicon matches or surpasses Boost Geometry's performance in bulk-loading. While both libraries utilize the same underlying packing strategy, Indexicon's lightweight node structure and avoidance of deep abstraction layers yield a slight edge in bulk-loading speeds. 
Indexicon's R-tree is significantly faster in dynamic insertions.
As expected, dynamic insertions are much more expensive than bulk-loading, so the large performance gap between the two structures gives Indexicon a big advantage over Boost in the overall construction cost.
Deletions follow the same trend. Indexicon is consistently faster than Boost in nearly all settings, with the only tie occurring on 2D MBBs from MIAMI. These results indicate that Indexomicon's architecture benefits both construction and index maintenance.

\begin{figure*}
    \centering
    \begin{subfigure}{\textwidth}
        \centering
        \includegraphics[width=\textwidth]{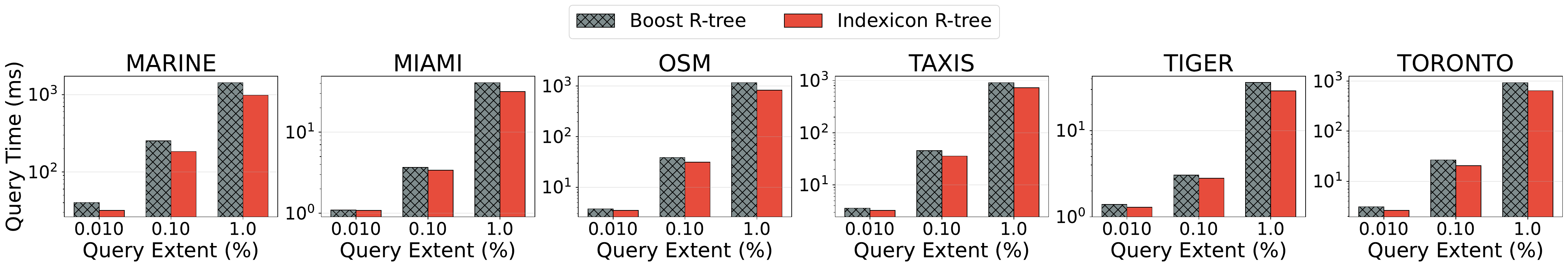}
        \caption{Range queries varying extent}
        \label{fig:rtree_2d_range}
    \end{subfigure}
    
    \vspace{1em}
    
    \begin{subfigure}{\textwidth}
        \centering
        \includegraphics[width=\textwidth, trim=0pt 0pt 0pt 73pt, clip]{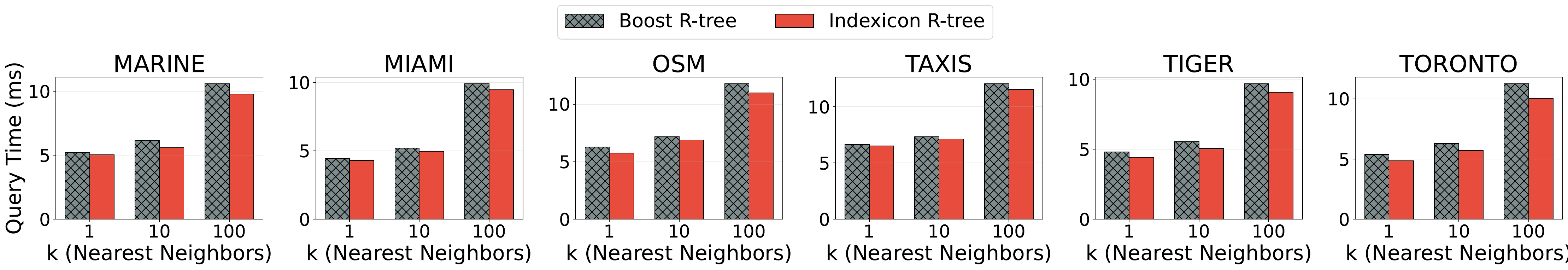}
        \caption{$k$NN queries varying $k$}
        \label{fig:rtree_2d_kNN}
    \end{subfigure}
    \caption{Query performance on 2D point datasets; default workload}
    \label{fig:rtree_2d_queries}
\end{figure*}

\stitle{Querying point datasets}.
Figure~\ref{fig:rtree_2d_queries} shows the performance of the two indices in spatial range and $k$-nearest neighbor ($k$NN) queries on 2D point data. Indexicon scales well as the query expands, outperforming Boost in all settings. For instance, on OSM's 1.0\% range queries, Indexicon executes the 1000 queries in 832.25 msecs, comfortably outperforming Boost's 1124.23 msecs execution time. This demonstrates that Indexicon's contiguous memory layouts and data locality directly translate to superior cache utilization during spatial scans. 
% Furthermore, Indexicon maintains a strict latency advantage for nearest neighbor searches. For example, with $k=100$ on TAXIS, Indexicon retrieves results in 11.07 msecs, outperforming Boost's 11.35 msecs.

% navigating its nodes much faster.

\begin{figure}
    \centering
    \begin{subfigure}{\columnwidth}
        \centering
        \includegraphics[width=\linewidth]{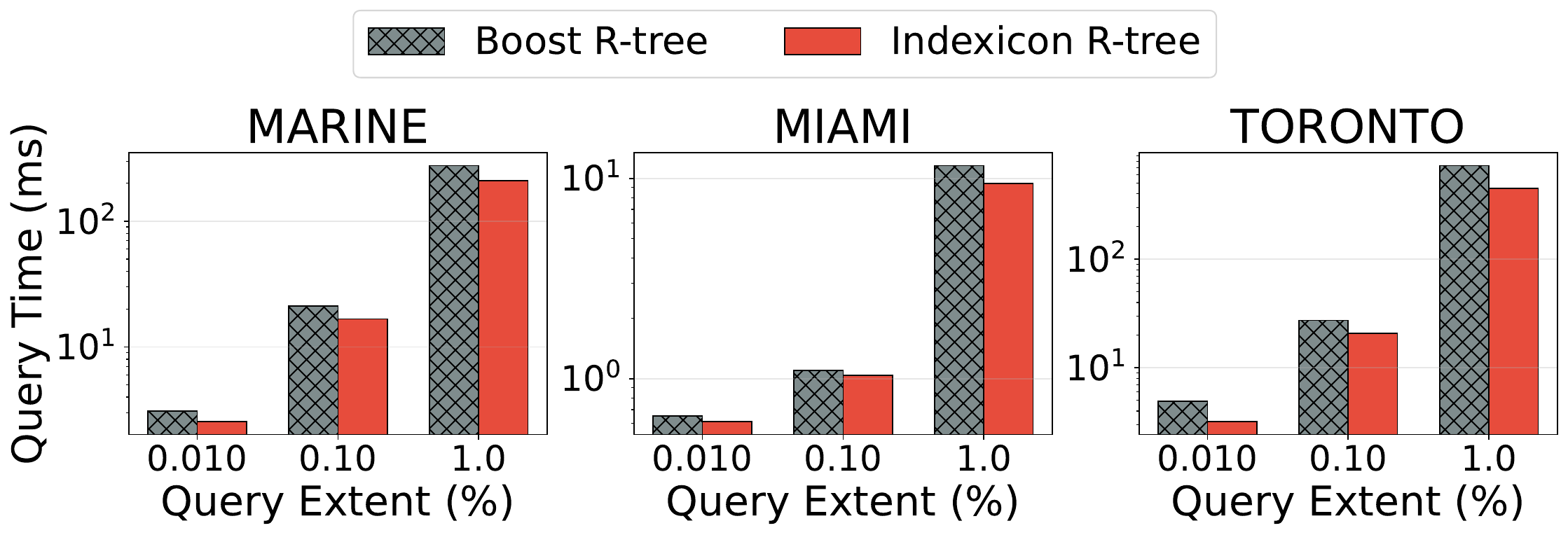}
        \caption{Range queries varying extent}
        \label{fig:rtree_3d_range}
    \end{subfigure}
    
    \vspace{1em}
    
    \begin{subfigure}{\columnwidth}
        \centering
        \includegraphics[width=\linewidth, trim=0pt 0pt 0pt 75pt, clip]{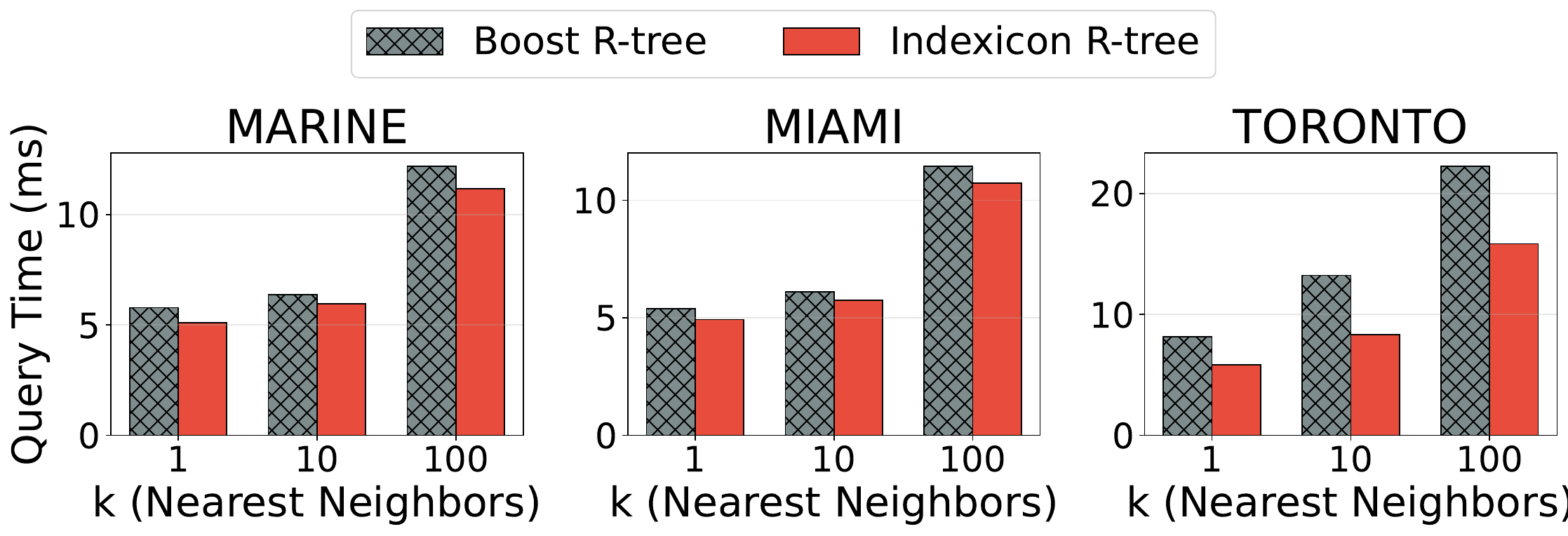}
        \caption{$k$NN queries varying $k$}
        \label{fig:rtree_3d_kNN}
    \end{subfigure}
    \caption{Query performance on 3D point datasets; default workload}
    \label{fig:rtree_3d_queries}
\end{figure}

% \paragraph{3D Point Datasets}
% We extend our evaluation to 3D spatial points. Table~\ref{tab:rtree_build_insert} details the construction and insertion latencies, showing that Indexicon provides a consistent advantage during bulk loading while outperforming Boost during dynamic updates. On MARINE, Indexicon completes dynamic insertions in 15.07 ss compared to Boost's 20.02 s, while on TORONTO, it finishes in 16.22 s versus Boost's 19.31 s.
Figure~\ref{fig:rtree_3d_queries} visualizes the execution times for both 3D range and $k$NN queries. The scaling advantages observed in 2D remain prominent in higher dimensions. On MARINE, Indexicon resolves 1000 queries of 1.0\% extent in 239.31 msecs versus Boost's 265.34 msecs. For $k$NN performance, Indexicon continues to retrieve neighbors faster. For example, for $k=100$ on TORONTO, Indexicon pulls the targets in 15.57 msecs, outperforming Boost's 21.03 msecs. 

\begin{figure}
    \centering
    \begin{subfigure}{\columnwidth}
        \centering
        \includegraphics[width=0.7\linewidth]{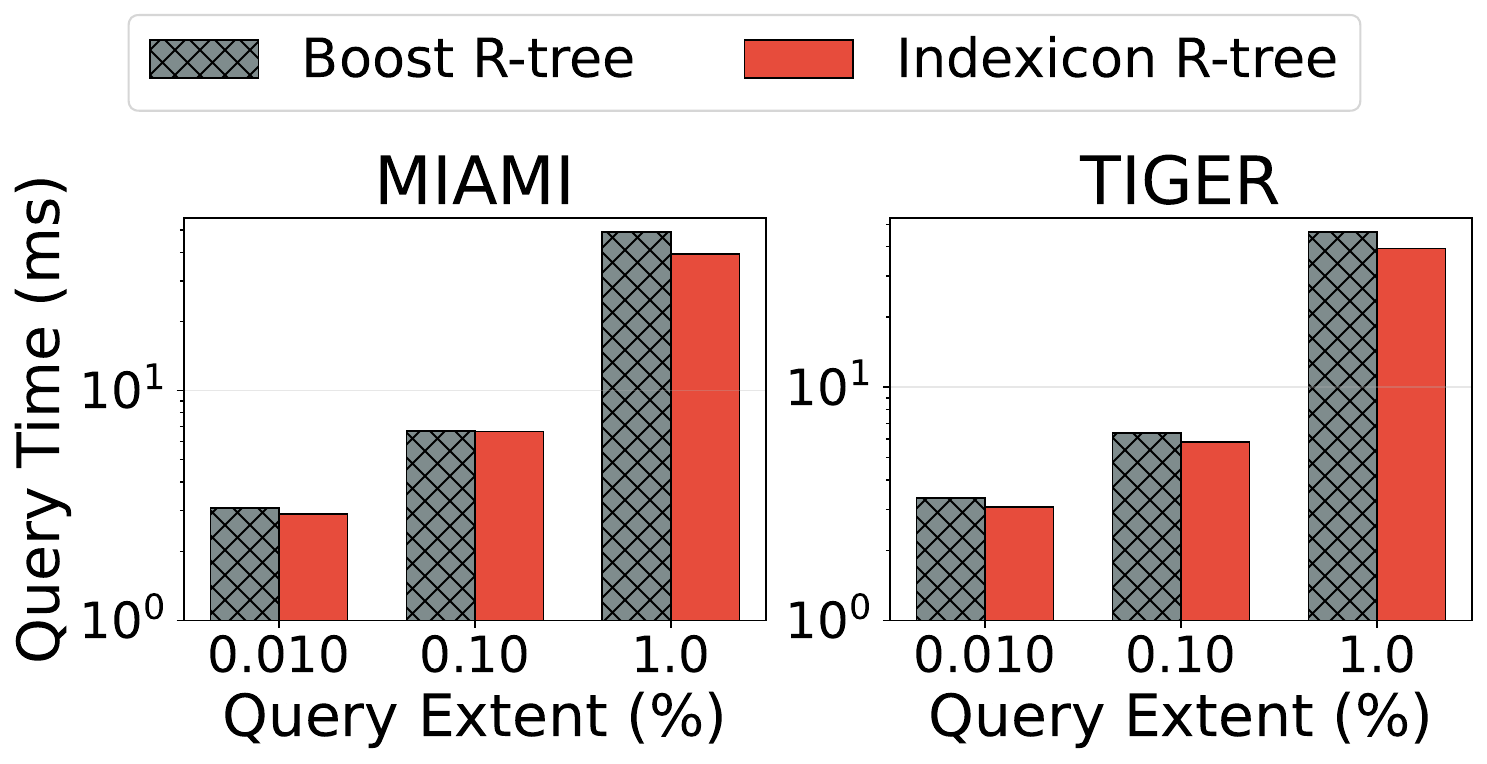}
        \caption{Range queries varying extent}
        \label{fig:rtree_2d_MBB_range}
    \end{subfigure}
    
    \vspace{1em}
    
    \begin{subfigure}{\columnwidth}
        \centering
        \includegraphics[width=0.7\linewidth, trim=0pt 0pt 0pt 75pt, clip]{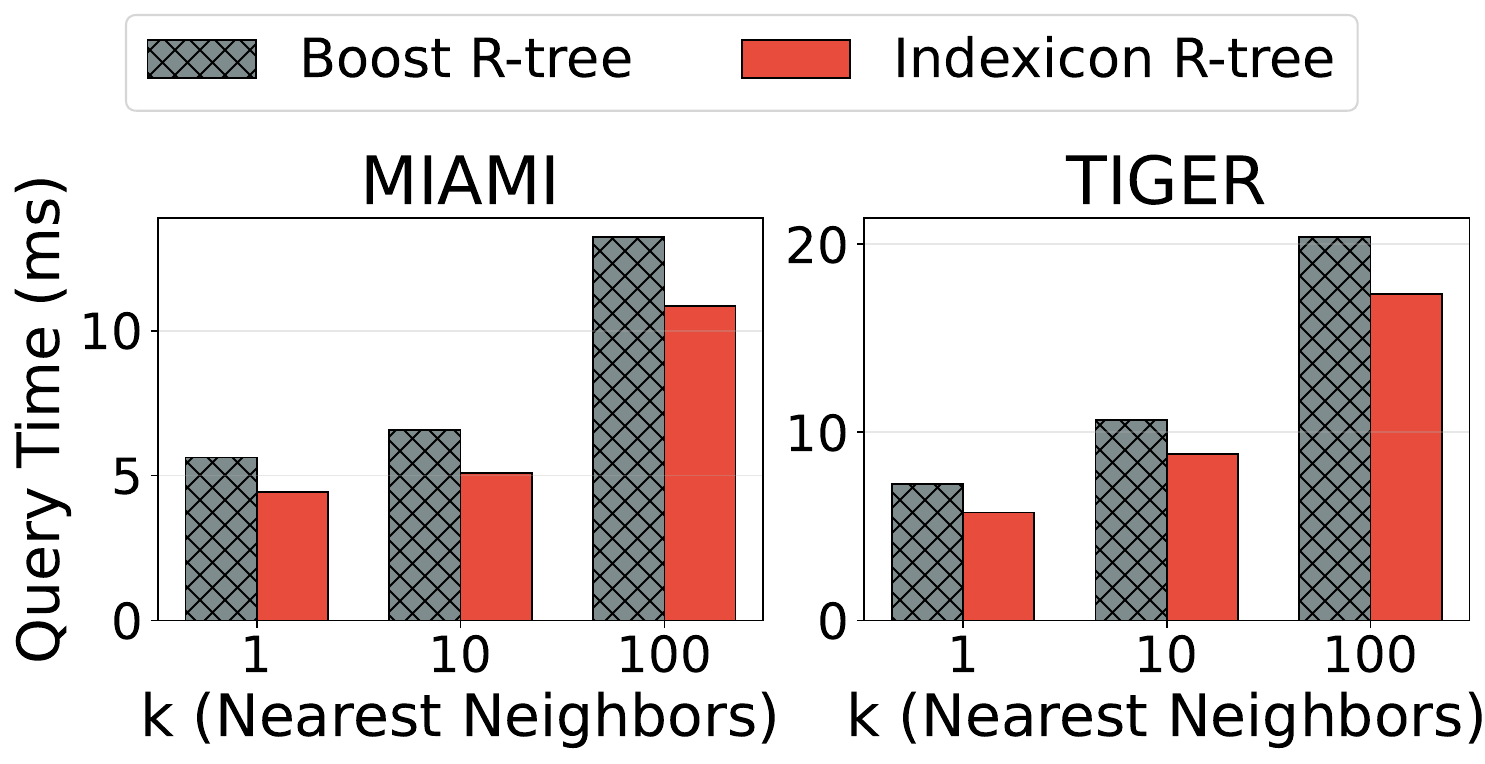}
        \caption{$k$NN queries varying $k$}
        \label{fig:rtree_2d_MBB_kNN}
    \end{subfigure}
    \caption{Query performance on 2D MBB datasets; default workload}
    \label{fig:rtree_2d_MBB_queries}
\end{figure}
\stitle{Querying MBB datasets}.
For 2D MBB data, Figure~\ref{fig:rtree_2d_MBB_queries} shows that Indexicon consistently outperforms Boost for both range and $k$NN queries. On TIGER, Indexicon evaluates the 1.0\% range queries in 39.38 msecs compared to Boost's 46.38 msecs, while on MIAMI, it requires 39.29 msecs compared to Boost's 48.90 msecs. The same trend appears for $k$NN search, where Indexicon benefits from its lightweight node representation. For example, for $k=100$, Indexicon takes 17.35 msecs on TIGER and 10.86 msecs on MIAMI, compared to Boost's 20.36 msecs and 13.23 msecs, respectively.

\begin{figure}[t]
    \centering

    % Common legend
    \includegraphics[width=0.5\columnwidth, trim=0pt 300pt 0pt 0pt, clip]{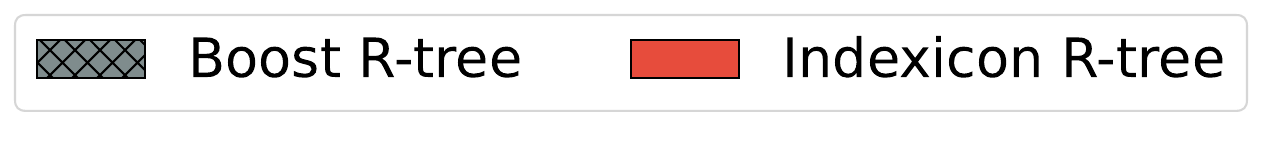}

    \vspace{0.5em}

    \begin{subfigure}{0.49\columnwidth}
        \centering
        \includegraphics[width=\linewidth, trim=90pt 0pt 50pt 75pt, clip]{figures/rtree_3d_mbr_mixed_delete_query.pdf}
        \caption{Range queries varying extent}
        \label{fig:rtree_3d_MBB_range}
    \end{subfigure}
    \hfill
    \begin{subfigure}{0.49\columnwidth}
        \centering
        \includegraphics[width=\linewidth, trim=90pt 0pt 50pt 75pt, clip]{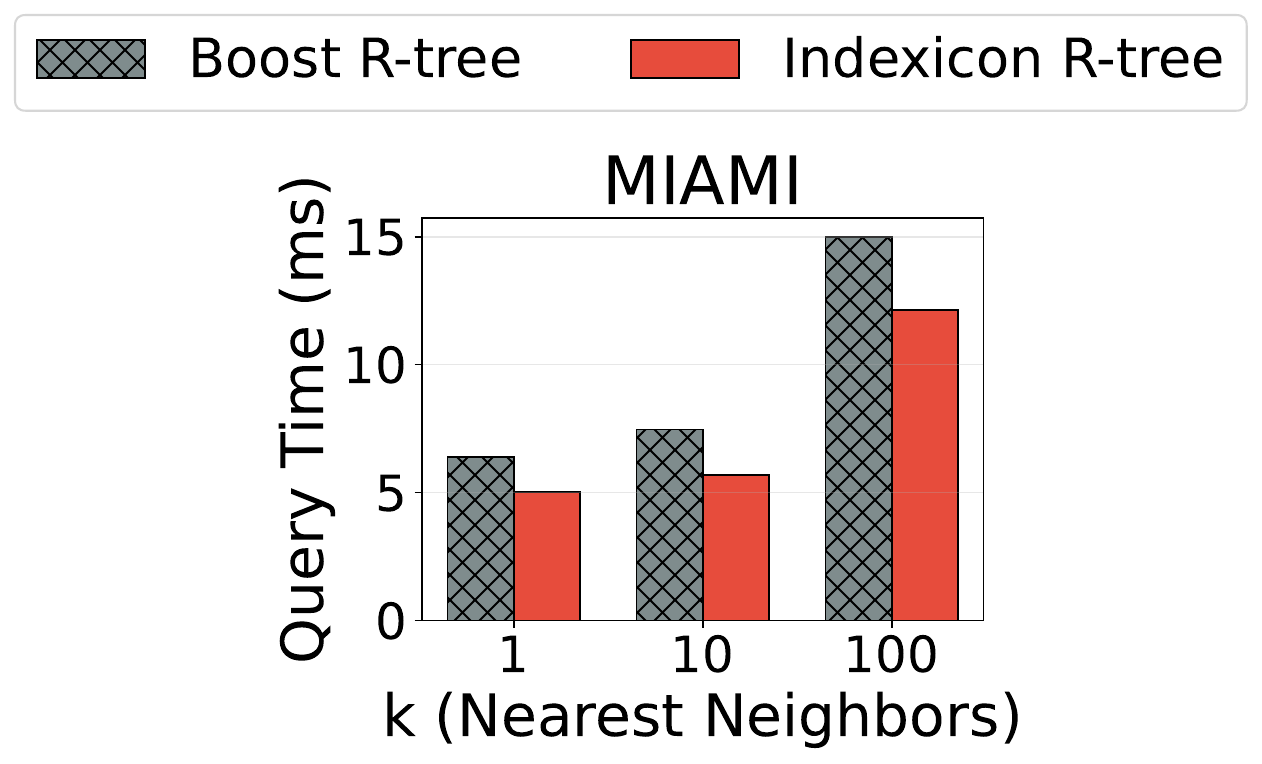}
        \caption{$k$NN queries varying $k$}
        \label{fig:rtree_3d_MBB_kNN}
    \end{subfigure}

    \caption{Query performance on 3D MBB data; default workload}
    \label{fig:rtree_3d_MBB_queries}
\end{figure}

For 3D MBB data, Figure~\ref{fig:rtree_3d_MBB_queries} shows an even larger performance gap since the additional dimension increases the cost of bounding-box comparisons. Indexicon remains faster across all query settings: for the 1.0\% range queries on MIAMI, it finishes in 56.78 msecs, while Boost requires 116.49 msecs, yielding roughly a 2$\times$ speedup. The advantage is also visible for smaller ranges, e.g., 7.68 msecs versus 11.90 msecs for the 0.10\% queries. For $k$NN queries, Indexicon again maintains a consistent lead; for $k=100$, it takes 12.14 msecs compared to Boost's 14.98 msecs. 
%These results validate Indexicon's implementation choices.
Overall, Indexicon exceeds search performance of Boost Geometry due to its structural quality, while offering a much simpler open-source codebase.

\subsection{Quad-tree and Oct-tree benchmark}
We now turn our focus to the Quad-tree and Oct-tree implementations. Before comparing Indexicon implementations to other libraries, we test the effect of the three partitioning strategies in Indexicon. 

\subsubsection{Effect of partitioning and parameters}

\begin{figure*}
    \centering
    \begin{subfigure}{\textwidth}
        \centering
        \includegraphics[width=\textwidth]{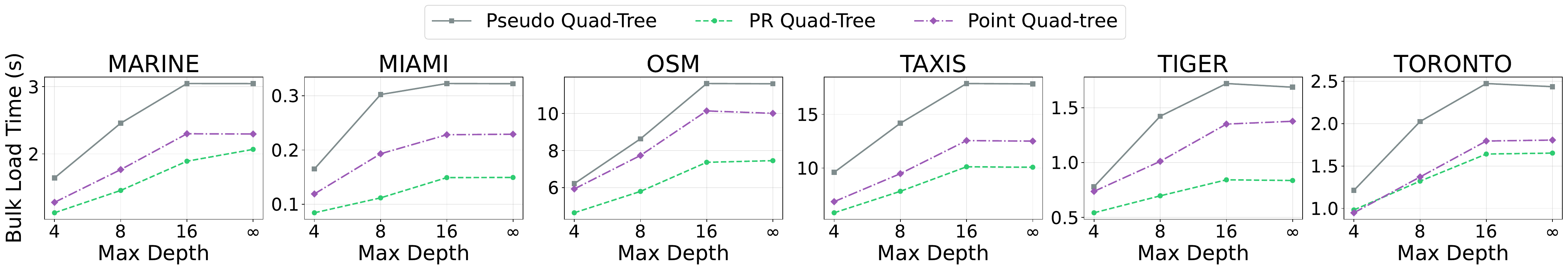}
        \caption{Bulk-loading}
        \label{fig:quadtree_depth_bulk}
    \end{subfigure}
    
    \vspace{1.5em}
    
    \begin{subfigure}{\textwidth}
        \centering
        \includegraphics[width=\textwidth, trim=0pt 0pt 0pt 75pt, clip]{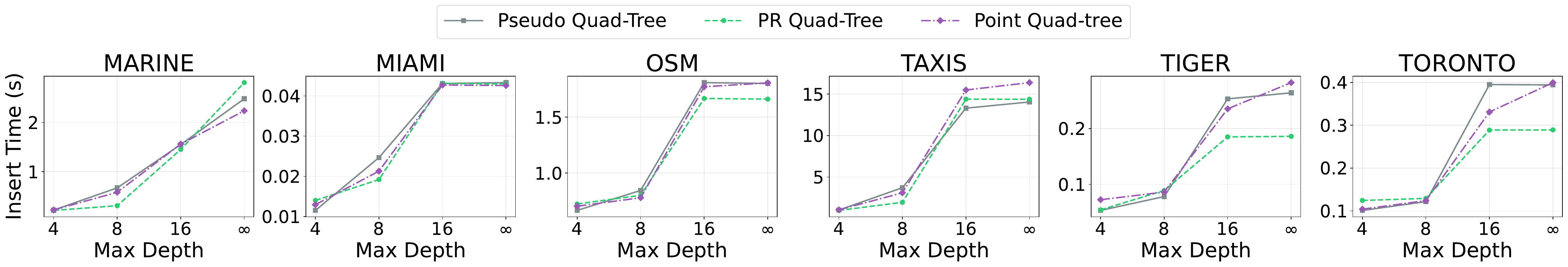}
        \caption{Dynamic insertions}
        \label{fig:quadtree_depth_insert}
    \end{subfigure}
    
    \vspace{1.5em}
    
    \begin{subfigure}{\textwidth}
        \centering
        \includegraphics[width=\textwidth, trim=0pt 0pt 0pt 75pt, clip]{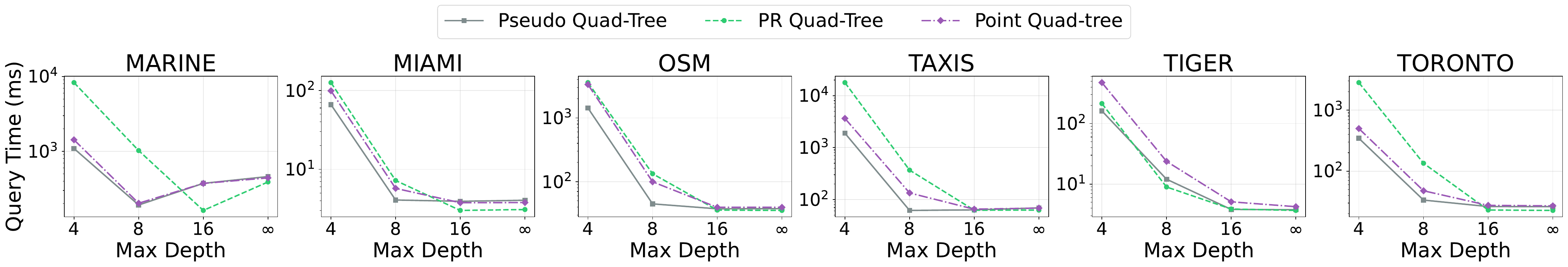}
        \caption{Range queries for 0.1\% extent}
        \label{fig:quadtree_depth_query}
    \end{subfigure}
    \caption{Impact of Quad-tree maximum depth constraints and splitting strategy construction and query performance; workload 80\% of each dataset bulk-loaded, 20\% insertions and 1000 queries}
    \label{fig:quadtree_depth_impact}
\end{figure*}

We evaluate three space-partitioning strategies under varying maximum tree depth constraints ($4$, $8$, $16$, and unlimited). The evaluated strategies (discussed in detail, in Section \ref{sec:indexes}) are: (1) the classic Point-Region (PR) Quad-tree which divides space into equal quadrants, (2) the Pseudo Quad-tree which splits at a synthetic coordinate consisting of the independent medians of both dimensions, and (3) a Point Quad-tree variant that splits on the physical median along the longest axis. We experiment with a workload which packs 80\% of the dataset into the index and inserts the rest.

Figure~\ref{fig:quadtree_depth_bulk} demonstrates the PR Quad-tree is the fastest structure to construct, while the point longest-axis strategy has an advantage over pseudo-median. For instance, at unlimited depth in OSM, point longest-axis completes bulk-loading in 10.01 secs compared to 11.61 secs for pseudo-median. This speedup is caused by the $O(1)$ axis-pruning selection which results in a single linear-time pass per split, cutting the split-finding in half relative to the dual-axis approach of pseudo-median. Dynamic update times shown in Figure~\ref{fig:quadtree_depth_insert} present no clear winner. In general, the PR Quad-tree is the fastest version for unlimited depth.

Figure~\ref{fig:quadtree_depth_query} reveals that imposing a strict maximum depth constraint heavily penalizes query performance regardless of the splitting strategy. A maximum depth of $4$ is a particularly bad choice, creating big leaves that effectively degrade spatial indexing into linear scan. Under such severely truncated depth bounds, the median-based strategies perform best, as their splits balance the data distribution, maximizing the partitioning utility of the few available tree levels. For example, pseudo-median executes the 0.1\% range queries on OSM in 1427.81 msecs, compared to 3341.39 msecs for point longest-axis and 3576.93 msecs for the PR Quad-tree. However, as the depth limit is relaxed (e.g., to 16) or left uncapped, this structural skew disappears, and the range query times of all methods converge. Since query performance smooths out when artificial truncation is removed, and the PR Quad-tree is superior in construction and insertion efficiency, we adopt the unconstrained PR Quad-tree as our default configuration for the remainder of this paper.

% 
% \nikos{again, group experiments by data type (point or MBB). Then, within each type split between 2D and 3D}
% 
% \paragraph{2D Point Datasets}
\begin{table}[h]
\centering
\scriptsize
\setlength{\tabcolsep}{2pt}
\caption{Quad-tree variants bulk-loading, insertion, and deletion times on 2D point datasets\eat{ for GEOS and Indexicon Quad-tree variants}; default workload}
\label{tab:geos_construct_delete}
\resizebox{\columnwidth}{!}{%
\begin{tabular}{lrrrrrrrrr}
\toprule
& \multicolumn{3}{c}{\textbf{Build (s)}} 
& \multicolumn{3}{c}{\textbf{Insert (s)}} 
& \multicolumn{3}{c}{\textbf{Delete (s)}} \\
\cmidrule(lr){2-4} \cmidrule(lr){5-7} \cmidrule(lr){8-10}
\textbf{Dataset} 
& \textbf{GEOS} & \textbf{PR} & \textbf{MX}
& \textbf{GEOS} & \textbf{PR} & \textbf{MX}
& \textbf{GEOS} & \textbf{PR} & \textbf{MX} \\
\midrule
MARINE  & 1.73 & 2.15 & 3.87  & 0.85 & 6.08  & 4.80  & 370.86 & 1.85 & 7.66 \\
MIAMI   & 0.36 & 0.16 & 0.31  & 0.24 & 0.23  & 0.25  & 0.20   & 0.10 & 0.13 \\
OSM     & -  & 7.25 & 13.82 & -  & 18.65 & 17.16 & - & 5.65 & 7.47 \\
TAXIS   & - & 10.11 & 19.93 & - & 17.90 & 17.26 & - & 5.82 & 6.67 \\
TIGER   & 0.86 & 0.82 & 1.68  & 0.63 & 1.76  & 1.70  & 73.32  & 0.65 & 0.87 \\
TORONTO & 1.44 & 1.58 & 2.94  & 0.81 & 2.53  & 2.40  & 72.78  & 0.91 & 0.96 \\
\bottomrule
\end{tabular}%
}
\end{table}
\begin{figure*}[t]
    \centering
    \begin{subfigure}{\textwidth}
        \centering
        \includegraphics[width=\textwidth]{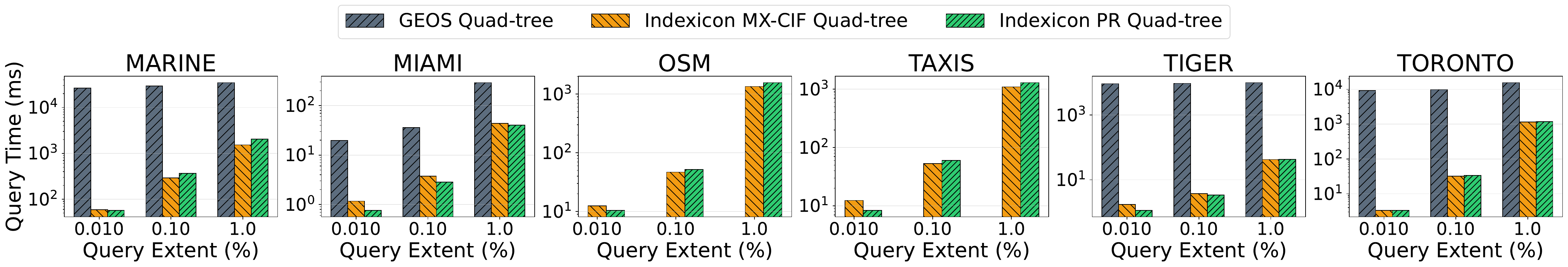}
        \caption{Range queries varying extent}
        \label{fig:geos_quadtree_range}
    \end{subfigure}

    \vspace{1em}

    \begin{subfigure}{\textwidth}
        \centering
        \includegraphics[width=\textwidth]{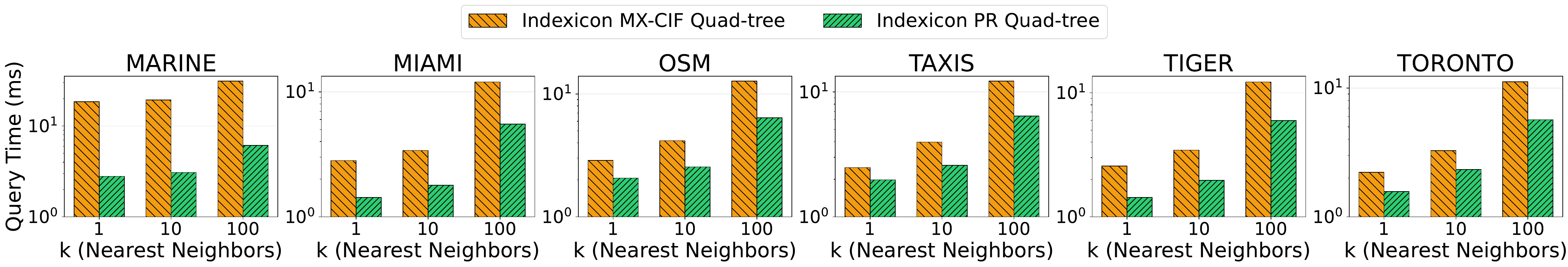}
        \caption{$k$NN queries varying $k$}
        \label{fig:geos_quadtree_knn}
    \end{subfigure}

    \caption{Query performance on 2D point datasets; default workload}
    \label{fig:geos_quadtree_query}
\end{figure*}

\subsubsection{Comparing Quad-tree and variants implementations}
\stitle{Point datasets}.
We compare the PR Quad-tree and MX-CIF of Indexicon against GEOS Quad-tree. 
% \nikos{Since we also have an MX-CIF Quad-tree implementation, this comparison seems unfair} \rapper{True but (i) Lei Chen uses it in his paper (ii) there arent other library implementations. private repos tend to have issues (I uppdated the Section 3.2)}
We subject both indices to the default workload. 
Table~\ref{tab:geos_construct_delete} reports construction and deletion performance for GEOS and Indexicon's Quad-trees on point data. GEOS exhibits fast insertions because its MX-CIF Quad-tree follows a different insertion policy from Indexicon's capacity-driven bucket Quad-trees. In Indexicon's PR Quad-tree, an overflowing leaf is split and its records are redistributed, which may create many additional nodes for dense or skewed point distributions. GEOS, instead, inserts items into the smallest existing or newly created quad that contains their envelope and allows nodes to store item lists directly. For point data, whose envelopes have zero area, GEOS also pads the envelope and uses a special insertion rule to avoid recursively creating deeper quads for zero-width envelopes. This limits subdivision and can lead to much shallower trees. For example, on MARINE, GEOS and Indexicon's PR Quad-tree consist of 10 and 25 levels, respectively. Within Indexicon, MX-CIF Quad-tree often inserts faster than the PR Quad-tree because it can stop earlier during insertion. Both variants split overflowing leaves and redistribute records, but MX-CIF may retain records at internal nodes when they lie on splitting axes. This reduces the number of leaf-level insertions and produces slightly fewer leaves. On OSM, after deletion, the PR Quad-tree contains 2.01M leaves compared with 1.99M leaves for MX-CIF. Thus, MX-CIF Quad-tree insertions often terminate slightly earlier.

However, GEOS's design can hurt deletions. Since many records may accumulate in node-local item lists, deleting an item may require searching through relatively large lists after the candidate node has been found. In contrast, Indexicon's capacity-based Quad-trees keep leaf buckets bounded, so deletion searches are usually over smaller local containers, although the tree may be deeper. This trade-off is visible on TIGER where deletion took 0.65 secs with Indexicon's PR Quad-tree, but 73.32 secs with GEOS. GEOS also failed to complete on OSM, while TAXIS was terminated due to excessive runtime.

Figure~\ref{fig:geos_quadtree_query} visualizes the range query execution times. The GEOS Quad-tree query operates as a primary filter, returning coarse candidate items whose spatial cells overlap the search window. To guarantee exact results, post-filtering is applied. This post-filtering, combined with the necessity to iterate through midline-straddling items, drastically degrades query performance. On the 1.0\% range queries for MARINE, the Indexicon PR Quad-tree resolves the query in 2,07 secs, outpacing GEOS, which struggles at 33,8 secs. Overall, the strict capacity constraints and bucketed leaves of the Indexicon's Quad-trees provide reliable construction and superior query and deletion throughput. Among the Indexicon variants, PR and MX-CIF Quad-trees exhibit comparable range-query performance, but PR has a clear advantage for $k$NN queries because it stores all records in leaf buckets and avoids the additional scans over internal-node straddle lists required by MX-CIF.

\begin{table}[h]
\centering
\footnotesize
\setlength{\tabcolsep}{4pt} 
% \caption{Bulk-loading and dynamic insertion times for the Indexicon Oct-tree versus the PCL Oct-tree on 3D point datasets.}
\caption{Oct-tree bulk-loading and dynamic insertion times on 3D point datasets; workload 80\% of each dataset bulk-loaded, 20\% insertions}
\label{tab:octtree_construct}

\begin{tabular}{lrrrr}
\toprule
& \multicolumn{2}{c}{\textbf{Build (s)}} & \multicolumn{2}{c}{\textbf{Insert (s)}} \\
\cmidrule(lr){2-3} \cmidrule(lr){4-5}
\textbf{Dataset} & \textbf{PCL} & \textbf{Indexicon} & \textbf{PCL} & \textbf{Indexicon} \\
\midrule
MARINE & 1.28 & 2.75 & 0.87 & 3.63 \\
TORONTO & 0.70 & 2.48 & 1.24 & 3.26 \\
\bottomrule
\end{tabular}%

\end{table}

\begin{figure}[h]
    \centering
    \includegraphics[width=0.7\columnwidth]{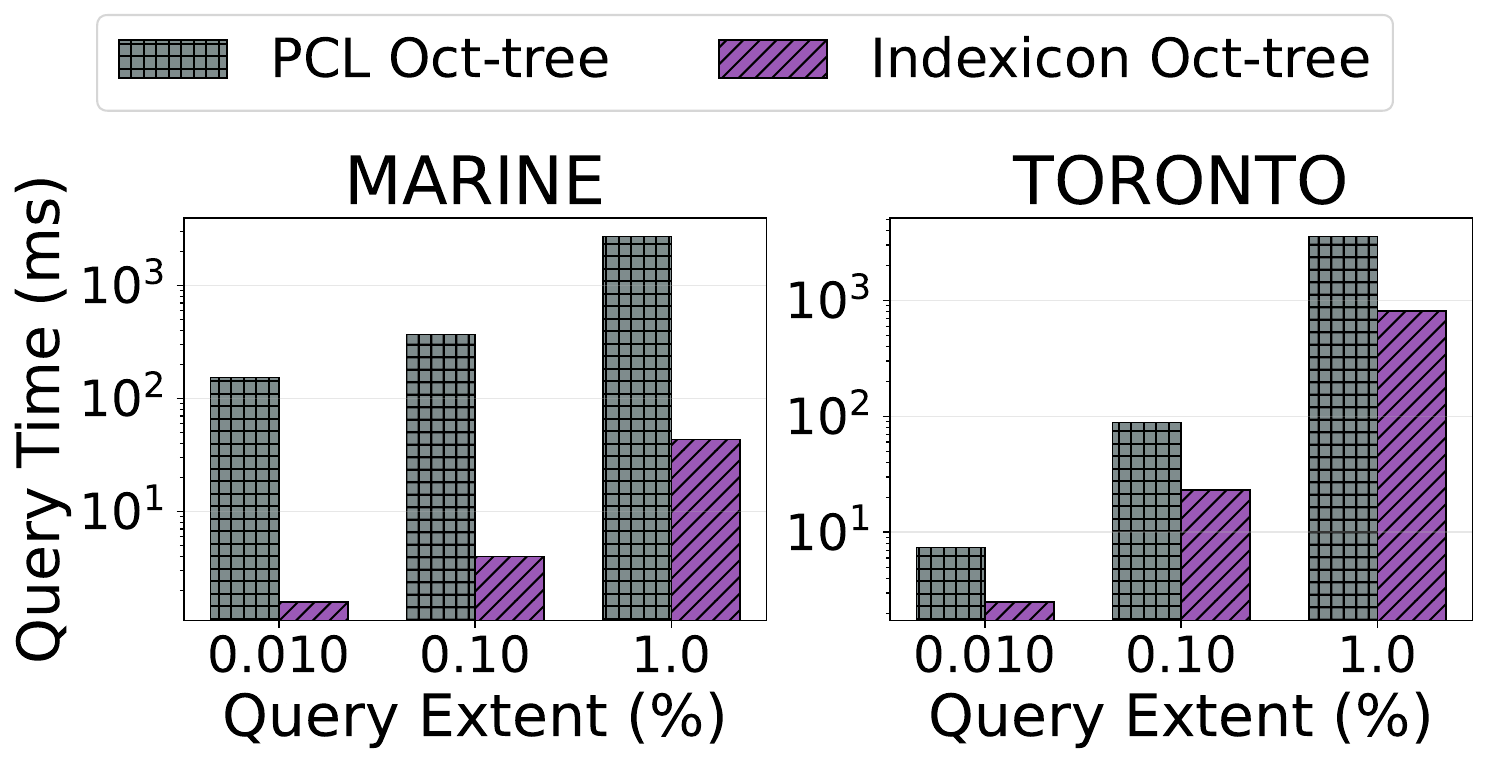}
    \caption{Query performance on 3D point datasets; workload 80\% of each dataset bulk-loaded, 20\% insertions}
    \label{fig:octtree_mixed_query}
\end{figure}

Table~\ref{tab:octtree_construct} reports construction and insertion times for the Indexicon Oct-tree and the PCL Oct-tree on 3D point datasets. PCL demonstrates faster build times because its Oct-tree relies on a resolution-driven, voxel-based architecture. Instead of storing full point records within the tree structure, its leaves store indices referencing the original point cloud array. This compact point-to-voxel mapping makes insertions highly efficient but hurt query performance. We enable PCL's dynamic-depth mode to enforce a leaf capacity limit, mimicking Indexicon's splitting behavior. However, this mechanism remains fundamentally bounded by PCL's maximum tree depth, which is dictated by the initial spatial resolution. 
In contrast, Indexicon's Oct-tree is a capacity-driven spatial index: overflowing leaves are split and their records are redistributed as part of the index structure. Thus, Indexicon performs more structural maintenance during construction and insertion, whereas PCL prioritizes a lightweight, index-only insertion path.
% \nikos{stop here, I think there is no need to mention 32-bit vs 64-bit. But you should mention whether the insertion strategy of PCL has a toll in query performance, i.e., whether query performance deteriorates because of this voxel-key based storage}
% Furthermore, PCL's restriction to 32-bit floats halves memory usage compared to Indexicon's full 64-bit double-precision calculations.
 % \nikos{this explanation does not make sense to me} \rapper{updated. an katalava kala, to PCL ftiaxnei ena 3D grid kai to xrisimopoiei san fylla tou Oct-tree}

In range queries (Figure~\ref{fig:octtree_mixed_query}), Indexicon has a large performance advantage over PCL. This is because PCL's lightweight insertion strategy takes a toll on query evaluation. First, PCL's index-based storage imposes a significant overhead due to indirect memory accessing; to evaluate exact bounds, candidates returned by a leaf node require random memory lookups into the external point array, leading to frequent CPU cache misses. Second, since PCL natively indexes data using 32-bit floats, to avoid false negatives, the query box must be artificially expanded in float space, and all candidates returned by PCL must be post-filtered against the exact 64-bit coordinates. By avoiding both indirect memory accesses and post-filtering, Indexicon achieves substantially higher query throughput. On the 1.0\% MARINE queries, the Indexicon Oct-tree is 63$\times$ faster than PCL. On TORONTO, the difference is smaller, around 4$\times$, but still significant.
% \nikos{does voxel-based storage also impose an overhead in queries (indirect accessing, cache misses, refinement, etc?) You should mention this if so}

\begin{table}[h]
\centering
\footnotesize
\setlength{\tabcolsep}{4pt} 
\caption{MX-CIF Quad-tree bulk-loading, dynamic insertion and deletion times on MBB datasets; default workload}
\label{tab:mxcif_construct}
\resizebox{\columnwidth}{!}{%
\begin{tabular}{lrrrrrr}
\toprule
& \multicolumn{2}{c}{\textbf{Build (s)}} & \multicolumn{2}{c}{\textbf{Insert (s)}} & \multicolumn{2}{c}{\textbf{Delete (s)}} \\
\cmidrule(lr){2-3} \cmidrule(lr){4-5} \cmidrule(lr){6-7}
\textbf{Dataset} & \textbf{GEOS} & \textbf{Indexicon} & \textbf{GEOS} & \textbf{Indexicon} & \textbf{GEOS} & \textbf{Indexicon} \\
\midrule
MIAMI & 0.22 & 0.29 & 0.17 & 0.16 & 0.28 & 1.34 \\
TIGER & 2.30 & 1.74 & 2.21 & 1.60 & 1.60 & 4.29 \\
\bottomrule
\end{tabular}%
}
\end{table}

\begin{figure}[h]
    \centering
    \includegraphics[width=0.80\columnwidth]{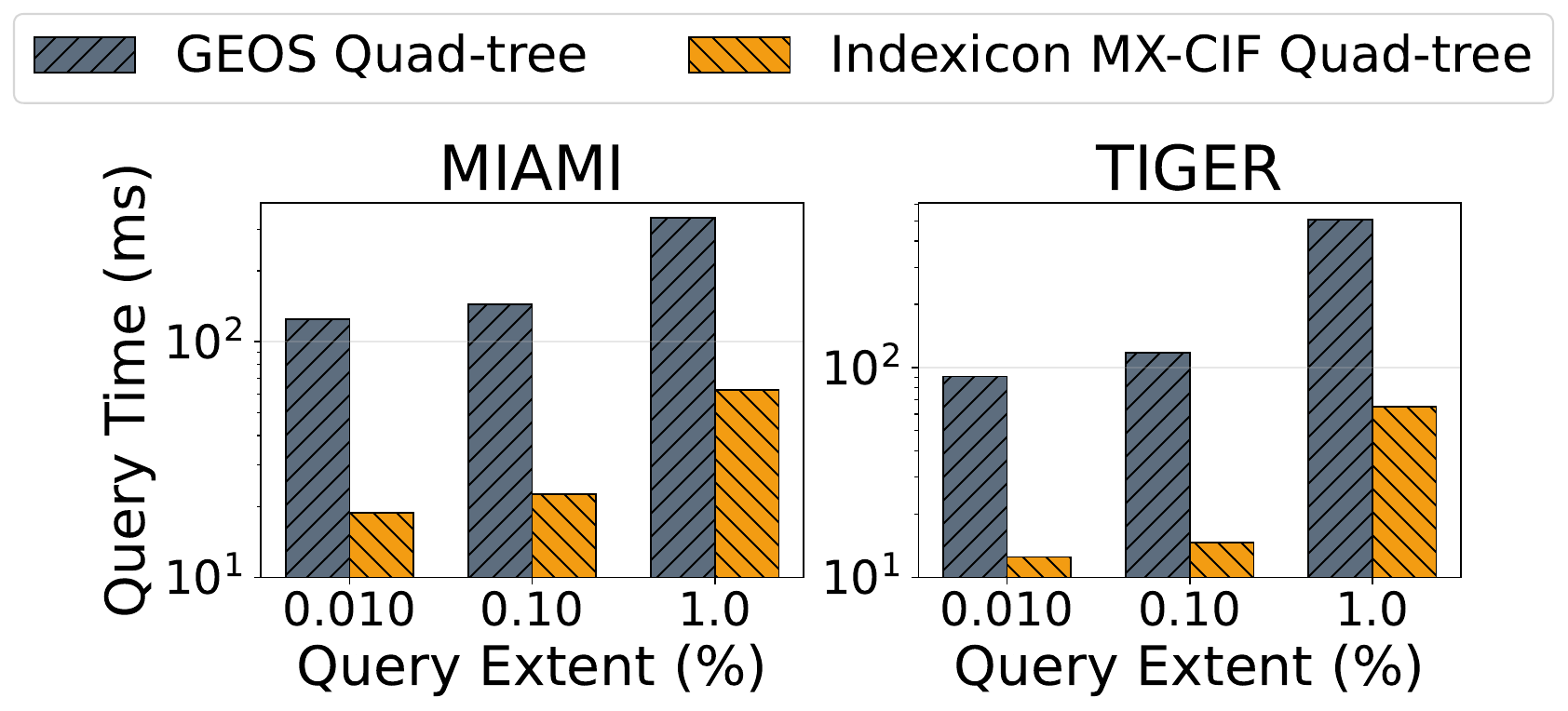}
    \caption{Query performance on MBB datasets, range queries varying extent; default workload \eat{\panos{shouldn't the legend have ``GEOS MX-CIF Quad-tree''?} \rapper{no, they call it quad-tree}}}
    \label{fig:mxcif_MBB_query}
\end{figure}

\stitle{MBB datasets}.
To evaluate indexing performance on MBB data, we compare the Indexicon MX-CIF Quad-tree against the GEOS MX-CIF Quad-tree under the default workload. Table~\ref{tab:mxcif_construct} reports construction and dynamic maintenance times for the MBB datasets. Indexicon is consistently faster during insertions, whereas GEOS is faster during deletions. For example, on TIGER, Indexicon completes insertions in 1.60 secs and deletions in 4.29 secs, while GEOS requires 2.21 secs for insertions but only 1.60 secs for deletions. This difference stems from the deletion policies of the two implementations. GEOS removes the item from a node, and prunes a child only if it is empty. In contrast, Indexicon may collapse four sibling leaves back into their parent if their combined contents fit within one bucket. This local consolidation adds deletion overhead, but keeps the tree layout more compact for subsequent queries.

As shown in Figure~\ref{fig:mxcif_MBB_query}, Indexicon's MX-CIF Quad-tree is significantly faster than GEOS in all cases. While GEOS needs a costly post-filtering step to eliminate false positives, Indexicon's MX-CIF Quad-tree natively evaluates exact bounds during traversal. Furthermore, our implementation utilizes a bucketed leaf architecture for non-straddling geometries, heavily optimizing cache locality during spatial scans. These results confirm that Indexicon's lean architecture scales efficiently for spatial extents.

\subsection{KD-tree benchmark}
Finally, we present our KD-tree benchmark. Before comparing Indexicon implementation against Nanoflann's KD-tree on point datasets, we investigate the effectiveness of the three KD-tree splitting strategies supported in Indexicon.

\begin{table}[h]
\centering
\footnotesize
\setlength{\tabcolsep}{4pt} 
\caption{KD-tree splitting strategies: bulk-loading and dynamic insertion times; workload 80\% of each dataset bulk-loaded, 20\% insertions}
\label{tab:kdtree_split}
\resizebox{\columnwidth}{!}{%
\begin{tabular}{lrrrrrr}
\toprule
& \multicolumn{3}{c}{\textbf{Build (s)}} & \multicolumn{3}{c}{\textbf{Insert (s)}} \\
\cmidrule(lr){2-4} \cmidrule(lr){5-7}
\textbf{Dataset} & \textbf{Round-robin} & \textbf{Adaptive} & \textbf{Longest-axis} & \textbf{Round-robin} & \textbf{Adaptive} & \textbf{Longest-axis} \\
\midrule
MIAMI & 0.24 & 0.27 & 0.21 & 0.15 & 0.15 & 0.14 \\
MARINE & 2.11 & 2.51 & 1.95 & 2.11 & 2.17 & 2.14 \\
OSM & 7.05 & 9.06 & 6.49 & 13.54 & 13.85 & 13.94 \\
TAXIS & 11.61 & 13.61 & 10.76 & 14.72 & 15.14 & 15.20 \\
TIGER & 1.14 & 1.48 & 1.06 & 1.60 & 1.62 & 1.57 \\
TORONTO & 1.72 & 2.19 & 1.59 & 1.83 & 1.87 & 1.83 \\
\bottomrule
\end{tabular}%
}
\end{table}

\begin{figure*}
    \centering
    \includegraphics[width=\textwidth]{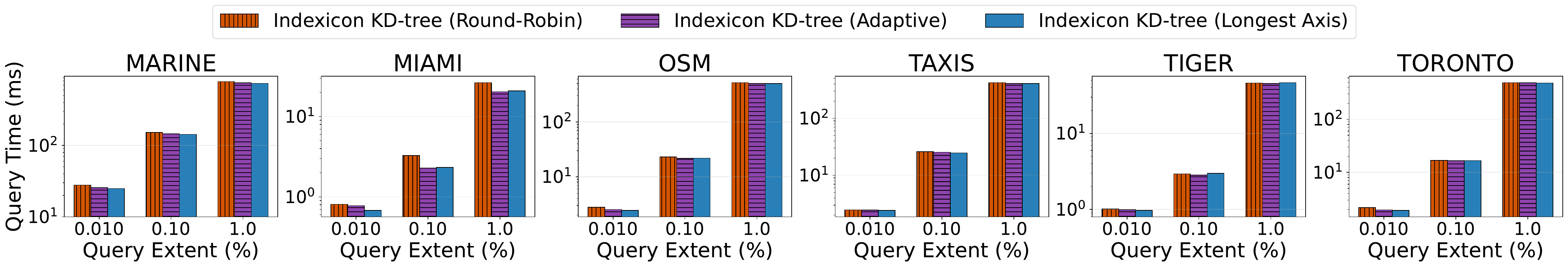}
    % \caption{Range query execution times for the Adaptive, Round-robin, and Longest-axis KD-tree strategies.}
    \caption{Impact of KD-tree splitting strategies on query performance for 2D point datasets; workload 80\% of each dataset bulk-loaded, 20\% insertions}
    \label{fig:kdtree_split_query}
\end{figure*}

\subsubsection{KD-tree splitting strategies}
We evaluate the three KD-tree splitting strategies in Indexicon: (1) the classic {\em round-robin} strategy, which cycles through the dimensions, (2) the \emph{adaptive} splitting strategy, which calculates the spatial spread of the elements and dynamically splits along the widest dimension, and (3) the \emph{longest-axis} approach, which determines the split by selecting each time the dimension with the largest bounding extent. We evaluate all variants using an 80\% bulk-loading and 20\% insertion workload. 

Table~\ref{tab:kdtree_split} details the construction phase. The adaptive strategy has the highest construction cost because it performs an additional scan of the records at every internal node to compute the minimum and maximum coordinate values per dimension before selecting the split axis. In contrast, both round-robin and longest-axis avoid this data scan. Round-robin selects the split dimension based on the tree depth, while longest-axis only compares the extents of the node's already maintained bounding region. Although round-robin has the cheapest split-axis selection rule, it is slower than longest-axis during bulk-loading (e.g., 7.05 secs vs 6.49 secs on OSM). This occurs because the dominant cost of bulk-loading is the in-place median partitioning step and not the split axis selection. By blindly cycling through dimensions, round-robin frequently forces the algorithm to partition along axes where the data is tightly clustered or contains many identical values, which degrades performance. Longest-axis avoids this by always splitting along the dimension with the widest spatial spread, making it the fastest bulk-loading strategy overall.

Despite the significant overhead that it imposes in the tree construction,
the adaptive strategy offers minimal query performance advantage, as shown in  Figure~\ref{fig:kdtree_split_query}. This happens because our KD-tree utilizes bucketed leaves. The adaptive and longest-axis splits were designed to prevent long, thin cells that degrade search performance. However, terminating the recursion early and storing elements in contiguous buckets naturally absorbs these geometric irregularities to a certain extent (see Figure~\ref{fig:partitioning_strategies}). Consequently, we select the longest-axis strategy as the default for our framework since it usually improves query performance (see MARINE, MIAMI, and OSM), with a minimal construction overhead.

% \paragraph{KD-tree 2D benchmark}

\begin{table}[t]
\centering
\footnotesize
\setlength{\tabcolsep}{3pt}
% \caption{KD-tree construction and update times.}
\caption{KD-tree bulk-loading, dynamic insertion and deletion times on point datasets; default workload}
\label{tab:kdtree_build_insert}
\begin{tabular}{lrrrrrr}
\toprule
& \multicolumn{3}{c}{\textbf{Nanoflann}} & \multicolumn{3}{c}{\textbf{Indexicon}} \\
\cmidrule(lr){2-4} \cmidrule(lr){5-7}
\textbf{Dataset} & \textbf{Build (s)} & \textbf{Insert (s)} & \textbf{Delete (s)} & \textbf{Build (s)} & \textbf{Insert (s)} & \textbf{Delete (s)} \\
\midrule
\multicolumn{7}{l}{\textbf{2D point}} \\
\midrule
MARINE & 5.04 & 2.61 & 0.18 & 2.01 & 2.30 & 1.11 \\
MIAMI & 0.20 & 0.26 & 0.02 & 0.20 & 0.14 & 0.09 \\
OSM & 7.41 & 16.35 & 1.17 & 6.49 & 13.95 & 5.07 \\
TAXIS & 21.99 & 20.45 & 1.28 & 10.89 & 15.02 & 5.65 \\
TIGER & 0.95 & 2.22 & 0.12 & 1.01 & 1.62 & 0.81 \\
TORONTO & 1.72 & 2.15 & 0.15 & 1.60 & 1.92 & 0.82 \\
\midrule
\multicolumn{7}{l}{\textbf{3D point}} \\
\midrule
MARINE & 3.91 & 2.18 & 0.18 & 2.05 & 2.34 & 1.01 \\
MIAMI & 0.22 & 0.29 & 0.02 & 0.22 & 0.16 & 0.10 \\
TORONTO & 2.00 & 2.58 & 0.15 & 1.73 & 2.00 & 0.90 \\
\bottomrule
\end{tabular}
\end{table}

\begin{figure*}[t]
    \centering
    \begin{subfigure}{\textwidth}
        \centering
        \includegraphics[width=\textwidth]{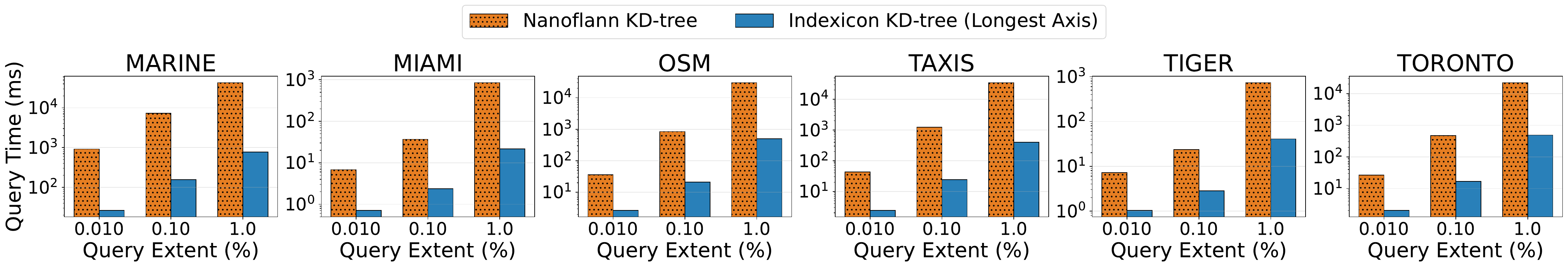}
        \caption{Range queries varying extent}
        \label{fig:kdtree_mixed_delete_query}
    \end{subfigure}
    
    \vspace{1em} 
    
    \begin{subfigure}{\textwidth}
        \centering
        \includegraphics[width=\textwidth, trim=0pt 0pt 0pt 75pt, clip]{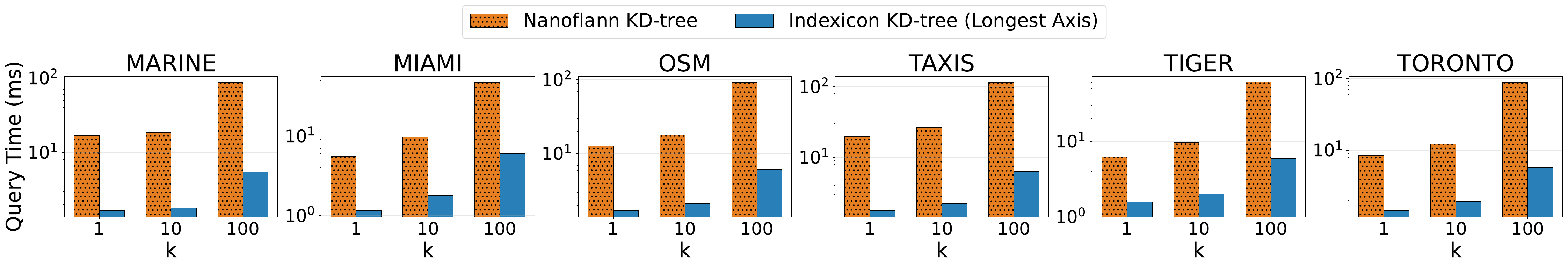}
        \caption{$k$NN queries varying $k$}
        \label{fig:kdtree_mixed_delete_$k$NN}
    \end{subfigure}
    \caption{Query performance on 2D point datasets; default workload}
    \label{fig:kdtree_mixed_queries_combined}
\end{figure*}

\subsubsection{Comparing KD-tree implementations}
We evaluate our KD-tree against Nanoflann, a popular KD-tree implementation on our default workload. Table~\ref{tab:kdtree_build_insert} details the latency of the structural modifications. It is critical to note that Nanoflann's dynamic adaptor performs multiple bulk-loads due to its underlying forest structure. Consequently, our KD-tree achieves superior construction times, with the only exception being TIGER, where Nanoflann has a slight edge. Conversely, the deletion times show a clear performance inversion with Nanoflann being drastically faster (e.g., 1.17 secs vs 5.07 secs on OSM). This behavior is an artifact of Nanoflann's lazy deletion strategy, which handles removals by flagging data as invalid (i.e., tombstones) without modifying the tree. While tombstones minimize structural mutation costs during writes, they shift the computational burden onto downstream reads as seen in our query experiments. 

Figure~\ref{fig:kdtree_mixed_queries_combined} demonstrates that our KD-tree outperforms Nanoflann by orders of magnitude across all settings in query times. For instance, on OSM's 1.0\% range query, KD-tree is roughly 60$\times$ faster than Nanoflann. This performance gap extends to $k$NN searches; for a $100$-NN search on OSM, our method registers a latency of 6.06 msecs compared to Nanoflann's 91.59 msecs. This is a consequence of the underlying structures and algorithmic constraints. Because our KD-tree remains a singular, cohesive tree optimized for continuous CPU cache prefetching, query traversals prune branches with maximum efficiency. In contrast, Nanoflann's queries iterate and merge results from its underlying forest while simultaneously spending clock cycles processing tombstones. This structural inefficiency is severely compounded during range searches which are further hurt by post-filtering.

\begin{figure}[t]
    \centering
    \begin{subfigure}{\columnwidth}
        \centering
        \includegraphics[width=0.9\columnwidth]{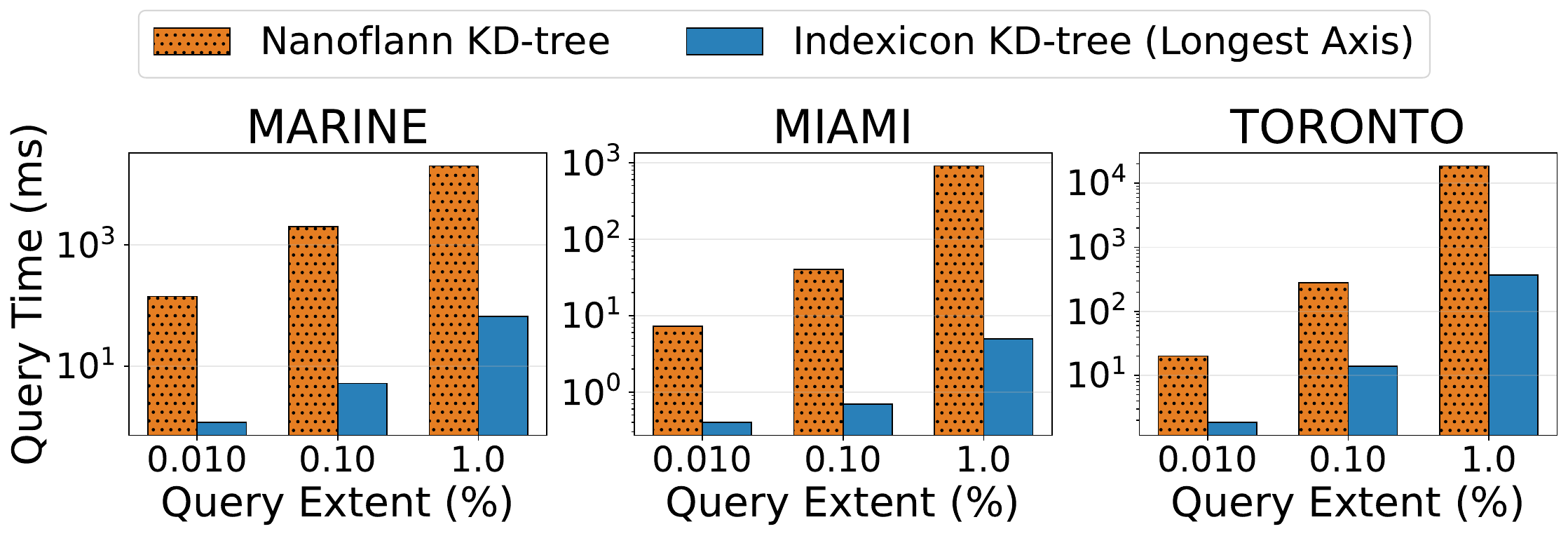}
        \caption{Range query varying extent}
        \label{fig:kdtree_3d_mixed_query}
    \end{subfigure}

    \vspace{2ex}
    \begin{subfigure}{\columnwidth}
        \centering
        \includegraphics[width=0.9\columnwidth, trim=0pt 0pt 0pt 75pt, clip]{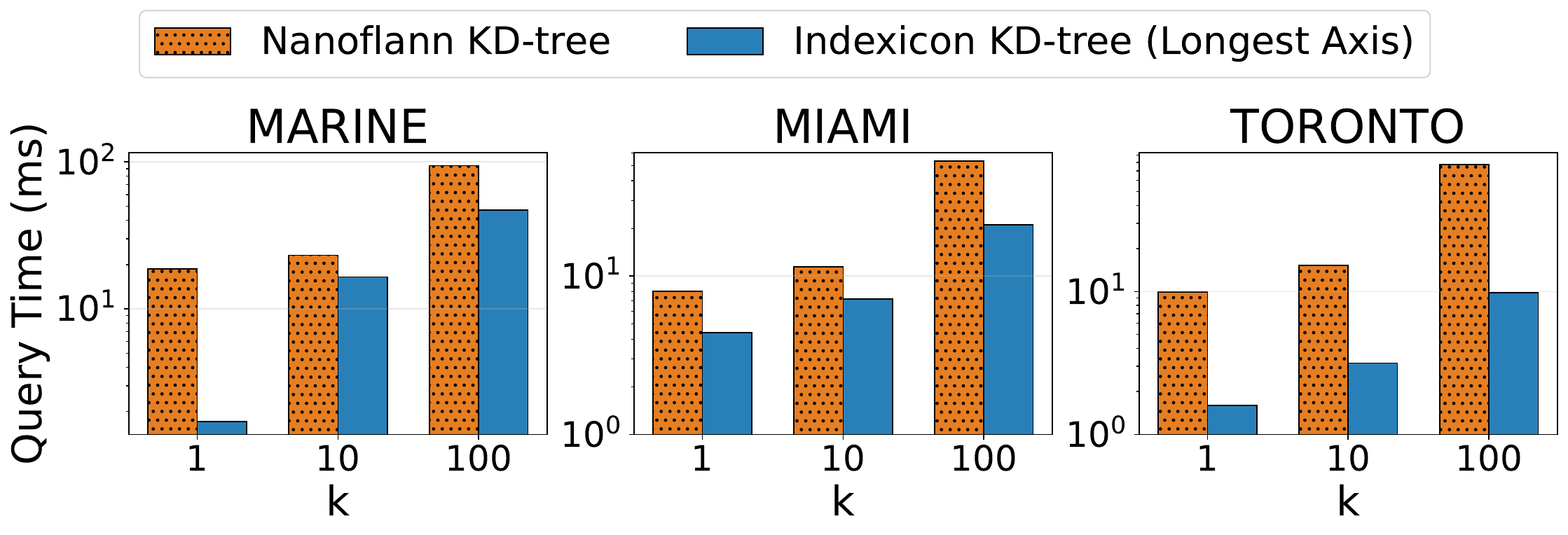}
        \caption{$k$NN queries varying $k$}
        \label{fig:kdtree_3d_mixed_$k$NN}
    \end{subfigure}
    \caption{Query performance on 3D point datasets; default workload}
    \label{fig:kdtree_3d_mixed_queries_combined}
\end{figure}

Next, we extend our evaluation to 3D point cloud data with the default workload. Table~\ref{tab:kdtree_build_insert} highlights the construction phase. Our KD-tree maintains its speed advantage during initial bulk-loading, completing the phase in 2.05 secs on MARINE compared to Nanoflann's 3.91 secs. During insertions, Nanoflann remains competitive, outperforming our KD-tree in MARINE but losing in the other datasets. As mentioned in the 2D case, Nanoflann's deletions remain fast due its tombstoning strategy.

Figure~\ref{fig:kdtree_3d_mixed_queries_combined} shows that the query latency gap remains in 3D space. Our KD-tree dominates Nanoflann across both range and $k$NN searches. On range queries, Nanoflann is orders of magnitude slower than Indexicon's KD-tree. For $k$NN queries, Nanoflann still underperforms, thus, showing that our KD-tree scales gracefully to 3D environments and the severe compounding effects of dynamic Nanoflann's architecture.

\section{Conclusions}\label{sec:conclusion}
We introduced Indexicon, a portable, open-source spatial indexing library engineered to bridge the gap between software simplicity and high-performance code for main-memory applications. By consolidating three foundational spatial structures, and their variants, into self-contained implementations, Indexicon systematically eliminates the fragmented APIs, heavy dependencies, and code complexity %\rapper{mipos einai ligo vary to code opacity?} 
that plague the current open-source spatial indexing ecosystem. Our extensive empirical evaluation across six real-world geographic datasets demonstrates that Indexicon consistently matches or outperforms competitive open-source indexing libraries and implementations  proving that structural flexibility does not require sacrificing execution efficiency.

\section{Future work}
\label{sec:future}
Although in its current state our library is fully deployable and addresses all three challenges outlined in the introduction, we are constantly evolving it along the following priority axes:
\begin{itemize}
% [noitemsep,topsep=2pt,parsep=0pt,partopsep=0pt,leftmargin=0.5cm]
    \item {\bf Improved support for updates.} Our KD-tree implementation currently does not support re-balancing under dynamic updates. We plan to investigate mechanisms that detect and repair imbalances caused by insertions or deletions. We also plan to extend the Oct-tree with additional 3D splitting strategies, including median-based variants analogous to those currently supported by our 2D Quad-tree implementations.
    
    \item {\bf Additional spatial indices.} We plan to expand Indexicon to include additional spatial indices (e.g., a 3D MX-CIF, KDB-Tree \cite{DBLP:conf/sigmod/Robinson81}, Waffle \cite{DBLP:journals/pvldb/MotiSP22}). Furthermore, recent studies \cite{DBLP:conf/icde/TsitsigkosLBMT21,MichalopoulosTBMT25,DBLP:conf/gis/MichalopoulosTB23,DBLP:journals/tkde/TsitsigkosBLMT24} show that grid-based structures can outperform tree-based ones in main memory. In the near future, we plan to implement uniform and adaptive 2D and 3D grids that support both point and MBB data. 
        
    \item {\bf Support for other query types.} In addition to range and nearest neighbor queries that we already support, we plan to include support for spatial (intersection and distance) joins in the implementations of the spatial indices.
    
    \item {\bf Multi-threaded search and updates.} We plan to include locking mechanisms to the spatial indices that enable high-throughput search and updates, while ensuring data consistency. For this, we plan to borrow ideas from our recent work in data-parallel and multi-threaded indexing \cite{bstree}.
\end{itemize}

\noindent Besides our own efforts to improve and expand Indexicon, we hope to engage the community in contributing to this project, offering improvements and extensions beyond the ones discussed above, constantly advancing Indexicon's open-source capabilities.

%% --- REFERENCES ---
\balance
\bibliographystyle{ACM-Reference-Format}
\bibliography{main} 

\end{document}